\documentclass{aa} 
\usepackage[varg]{txfonts}
\usepackage{graphicx}
\usepackage{graphics}
\usepackage{amssymb}
\usepackage{amsmath}
\usepackage[T1]{fontenc}
\usepackage[utf8]{inputenc}
\usepackage[english]{babel}
\usepackage{color}
\usepackage{natbib}
\bibpunct{(}{)}{;}{a}{}{,} 
\usepackage[switch]{lineno}
\usepackage{multicol}

\newcommand{\astrosat}{{\it AstroSat}}
\newcommand{\swift}{{\it Swift}}
\newcommand{\nustar}{{NuSTAR}}
\newcommand{\fermi}{{\it Fermi}-LAT}

\newcommand{\pd}[1]{\, \partial #1 \,}
\newcommand{\td}[1]{\, {\rm d} #1 \,}
\newcommand{\intl}{\int\limits}

\newcommand{\SF}[3]{\; H\left[ #1;\, #2,\, #3 \right]}  
\newcommand{\g}{\ensuremath{\gamma}}

\newcommand{\p}{^{\prime}}

\newcommand{\E}[1]{\times 10^{#1}}
\newcommand{\lhpi}{\textit{LH$\pi$}}
\newcommand{\lhp}{\textit{LHp}}
\newcommand{\SSC}{\textit{SSC}}
\newcommand{\epshock}{\textit{e-p-shock}}
\newcommand{\eTeV}{\textit{extreme}-TeV}
\newcommand{\esyn}{\textit{extreme}-Syn}
\newcommand{\onehale}{\textsc{OneHaLe}}
\begin{document}
%
%
%
\title{The variety of extreme blazars in the \astrosat\ view}
%
\author{
P.~Goswami$^{1,2}$, M.~Zacharias$^{3,2}$, A.~Zech$^4$, S.~Chandra$^{2,5}$, M.~Boettcher$^2$, I.~Sushch$^{2,6}$
}
%

\institute{
$^1$Université Paris Cité, CNRS, Astroparticule et Cosmologie, F-75013 Paris, France, goswami@apc.in2p3.fr \\
$^2$Centre for Space Research, North-West University, Potchefstroom, 2520, South Africa \\
$^3$Landessternwarte, Universit\"{a}t Heidelberg, K\"{o}nigstuhl 12, 69117 Heidelberg, Germany, m.zacharias@lsw.uni-heidelberg.de \\
$^4$Laboratoire Univers et Théories, Observatoire de Paris, Université PSL, CNRS, Université de Paris, 92190 Meudon, France \\
$^5$South African Astronomical Observatory, Observatory Road, Cape Town, 7925, South Africa \\
$^6$Astronomical Observatory of Ivan Franko National University of Lviv, Kyryla i Methodia 8, 79005 Lviv, Ukraine
}
%
\date{Received ? / accepted ? }

\abstract{
Among the blazar class, extreme blazars have exceptionally hard intrinsic X-ray/TeV spectra, and extreme peak energies in their spectral energy distribution (SED). Observational evidence suggests that the non-thermal emission from extreme blazars is typically non-variable. All these unique features present a challenging case for the blazar emission models, especially for the sources with hard TeV spectra.} 
{We aim to explore the X-ray and GeV observational features of a variety of extreme blazars, including  \eTeV, {\it extreme}-Synchrotron and regular High-frequency-peaked BL Lac objects (HBLs). Furthermore, we aim to test the applicability of various blazar emission models that could explain the very hard TeV spectra.} 
{We conduct a detailed spectral analysis of X-ray data collected with \astrosat\ and \swift-XRT, along with quasi-simultaneous \g-ray data from \fermi, for five sources; 1ES\,0120+340, RGB\,J0710+591, 1ES\,1101-232, 1ES\,1741+196 and 1ES\,2322-409. To model the SEDs, we employ three approaches: 1) a steady-state one-zone synchrotron-self-Compton (SSC) code, 2) another leptonic scenario of co-accelerated electrons and protons on multiple shocks, applied only on the \eTeV sources and 3) a one-zone hadro-leptonic (\onehale) code. The hadro-leptonic code is used twice to explain the \g-ray emission process:  proton synchrotron and synchrotron emission of secondary pairs.

} 
%
{Our X-ray analysis provides well-constrained estimates of the synchrotron peak energies for both 1ES0120+340 and 1ES1741+196. These findings categorize them as {\it extreme}-Synchrotron sources, as they consistently exhibit peak energies above 1\,keV in different flux states.
The multi-epoch X-ray and GeV data reveal spectral and flux variabilities in RGB\,J0710+591 and 1ES\,1741+196, even on time scales of days to weeks. As anticipated, the one-zone SSC model adequately reproduces the SEDs of regular HBLs but encounters difficulties in explaining the hardest TeV emission. Hadronic models offer a reasonable fit to the hard TeV spectrum, though with the trade-off of requiring extreme jet powers. On the other hand, the lepto-hadronic scenario faces additional challenges in fitting the GeV spectra of {\it extreme}-TeV sources. Finally, the e-p co-acceleration scenario naturally accounts for the observed hard electron distributions and effectively matches the hardest TeV spectrum of RGB\,J0710+591 and 1ES\,1101-232. } 
{} 
%
%
\titlerunning{\astrosat~ view on extreme blazars} 
\authorrunning{P. Goswami et al.} 
\maketitle



\section{Introduction}\label{sec:introduction}


Blazars are a subclass of active galactic nuclei (AGN) that emit non-thermal, strongly polarized and variable continuum emission from a jet of relativistic plasma directed along or close to the line of sight \citep{RogerDBlandford_1978, 1995PASP..107..803U}. The broadband spectral energy distribution (SED) of blazars displays two broad humps: the low energy emission (peaking in the sub-mm to soft X-ray range) is commonly ascribed to synchrotron emission from relativistic electrons in the jet, while the origin of the high energy emission component (peaking at MeV to TeV energies) continues to be a subject of discussion with various proposed solutions \citep{2010ApJ...716...30A}. Two viable scenarios, leptonic and hadronic, are widely used to explain the high-energy emission. Leptonic models propose that the high-energy emission comes from inverse Compton scattering of low-energy seed photons by ultra-relativistic leptons, either from the synchrotron radiation field in the emission region \citep[Synchrotron Self-Compton, SSC, e.g.,][]{1989ApJ...340..181G, 1996ApJ...461..657B} or from photons originating external to the emission region \citep[External Compton, EC, e.g.,][]{1992A&A...256L..27D, 1994ApJ...421..153S}. Hadronic models, on the other hand, assume that the high-energy emission originates from accelerated ultrarelativistic protons in the jet, through the proton synchrotron mechanism or secondary emission from particles such as electron-positron pairs or muons produced in $p\gamma$ interactions \citep{1992A&A...253L..21M, 2001APh....15..121M, Boettcher13}.


Extreme blazars (or eHBLs), a peculiar class of high-energy peaked blazars, pose a significant challenge to conventional blazar models due to their unique spectral characteristics (\cite{2001A&A...371..512C, 2020NatAs...4..124B} for a recent review). The eHBLs are typically characterized by an unusually hard intrinsic spectrum (photon index, $\Gamma \sim 1.5-1.9$) in both their X-ray and very-high-energy (VHE) gamma-ray emissions, and their SED peaks reaching up to 1-10\,keV (typically $>$ 1 keV) 
in the synchrotron component, and a few TeV ($>$1-10 TeV) in the high-energy component consistently in different flux states. It is worth noting that the extreme properties observed in these two energy bands do not always coexist and the correlation between the two extreme properties remains unknown \citep{2019MNRAS.486.1741F, Costamante+2018}. Two types of blazars are recognized as being extreme: extreme synchrotron blazars ({\it extreme}-Syn, e.g., 1ES~0033+595, 1ES~0120+340) and extreme TeV blazars ({\it extreme}-TeV, e.g., 1ES~0229+200, 1ES~0347-121). However, these are distinct from transiting high-synchrotron-peaked blazars (HBLs) --- such as Mkn 421, Mkn 501, 1ES\,1426+428, 1ES~2344+514, and 1ES\,1959+650 --- which only exhibit extreme behaviour during strong flares. In contrast, eHBLs are not known to show such strong flares and exhibit persistently extreme behaviour in different flux states.

Due to the low flux detectability and limited observational range in the X-ray and very high energy (VHE) gamma-ray bands, there is still considerable uncertainty in locating the SED peak positions of extreme sources, and only a few have been identified thus far. Several sources have been classified as {\it extreme}-Syn sources or potential sources based on BeppoSAX observations \citep{2001A&A...371..512C}, while a few have been confirmed by \cite{Costamante+2018} through precise localization of the synchrotron peak using joint XRT-NuSTAR observations. In the case of VHE gamma-rays, the number of confirmed extreme sources is more than ten \citep{2020NatAs...4..124B, 2020ApJS..247...16A}. Among the observed eHBLs, 1ES\,0229+200 is the best example that displays high peak frequencies in both X-rays and VHE gamma-rays. Hence, the source is of great importance for jet physics, as well as constraining important cosmological quantities such as the extragalactic background light and the intergalactic magnetic field \citep{2007A&A...475L...9A, 2010MNRAS.406L..70T, 2015MNRAS.451..611B}. 

 
Unlike most TeV blazars, which exhibit significant fluctuations and flares, eHBLs appear to display relatively stable emissions. Despite the lack of strong flares or high flux activities on minute timescales at any wavelength, recent observations have indicated that moderate variability can be present in some eHBLs. For example, in X-rays, 1ES\,1101-232 showed a variation of about 30\% in flux and corresponding spectral variability \citep{2000A&A...357..429W}. The TeV lightcurve of 1ES\,1218+304 exhibited rapid TeV variability over a few days, reaching approximately 20\% of the Crab flux \citep{2010ApJ...709L.163A}, while 1ES\,0229+200 displayed moderate variations in TeV on yearly timescales \citep{2014ApJ...782...13A}. These findings contradict the idea that the absence of variability is a universal feature of eHBLs.


A large variety of models under leptonic and hadronic scenarios have been proposed to explain the extreme emission. While a simple synchrotron self-Compton (SSC) model provides a good explanation for regular blazars and can also account for the extreme synchrotron emission observed in some sources, interpreting the extremely hard TeV spectrum within a purely leptonic SSC framework is challenging. This often requires a large value of the minimum Lorentz factor ($\gamma_{min} \sim 10^4-10^5$) \citep{2006MNRAS.368L..52K, 2011A&A...534A.130K} or hard particle spectra, as well as a very weak magnetic field ($B \leq 1 mG$) \citep{Costamante+2018}. The limitations of the one-zone SSC model are widely discussed by \cite{cerruti+15},  \cite{2022MNRAS.512.1557A}, and \cite{2020NatAs...4..124B}. Alternative approaches have been proposed to explain extreme TeV emission within the leptonic framework. For instance, an external Compton scenario, which involves the Compton upscattering of Cosmic Microwave Background (CMB) photons in the extended kiloparsec (kpc)-scale jet \citep[1ES\,1101-232,][]{2008ApJ...679L...9B, 2012MNRAS.424.2173Y} and photons from the broad-line region \citep{2011ApJ...740...64L}, or the internal gamma-ray absorption scenario. \citep{2008MNRAS.387.1206A, 2011ApJ...738..157Z}. However, the short-term variability detected in some sources seems incompatible with such a kpc-scale-jet scenario.
Another approach involves taking into account adiabatic losses or a Maxwellian-type electron distribution in a stochastic acceleration model, which leads to a very hard TeV spectrum \citep[1ES\,0229+200,][]{2011ApJ...740...64L}.

In a recent work, \cite{Zech2021} proposed a feasible solution to address the issues associated with the pure SSC model by providing more natural explanations for the requirement of large values of 
the minimum electron Lorentz factor and low magnetization. This model is an extension of the standard SSC theory and assumes that both electron and proton populations are co-accelerated in relativistic internal or recollimation shocks. Possible energy transfer mechanisms can naturally result in a very high value of $\gamma_{min}$. The model considers different shock and recollimation scenarios that can explain extreme 
($\Gamma_{VHE} \sim 1.7-1.9$, e.g., RGB\,J0710+591, 1ES\,1101-232) to very extreme ($\Gamma_{VHE} \sim 1.5$, e.g., 1ES\,0229+200) VHE $\gamma$-ray spectra and apparently require recollimation at more than a single shock to produce the 
hardest VHE spectra. Further, an adaptation of the \cite{Zech2021} model was proposed by \cite{Tavecchio2022} where the extremely hard TeV emission is explained by a combination of recollimation and stochastic acceleration.


On the other hand, different flavours of hadronic models (proton synchrotron and secondary cascades produced in $p\gamma$ interactions) have advantages over a standard leptonic model and somewhat relax the requirements for extreme parameter values. For instance, the lepto-hadronic solution suggested by \cite{cerruti+15} has effectively replicated an extremely hard TeV spectrum, albeit with a demand for hard injection functions. Another lepto-hadronic approach recently explored by \cite{2022MNRAS.512.1557A} suggested that the extreme emission is coming from photo-hadronic interactions in a blob close to the AGN core and by SSC and external inverse Compton-processes in an outer blob. Nevertheless, (lepto-)hadronic models, in general, demand very high proton power, 
sometimes with super-Eddington values in the cases of extreme TeV sources. 
\cite{refId0} devised a one-zone model based on hadronuclear ($pp$) interactions, which circumvents extreme jet-power requirements. 

This paper reports on recent observations carried out using \astrosat\ and \fermi\ of five sources: 1ES\,0120+340 (redshift $z=0.272$), RGB\,J0710+591 ($z=0.125$), 1ES\,1101-232 ($z=0.186$), 1ES\,1741+196 ($z=0.084$), and 1ES\,2322-409 ($z=0.1736$), each displaying a unique range of spectral characteristics. Among these, RGB\,J0710+591 and 1ES\,1101-232 are well-known for being {\it extreme}-TeV sources with hard intrinsic TeV spectra. Although TeV data is unavailable for 1ES\,0120+340, it presents itself as a potential \eTeV\ candidate with hard X-ray and GeV spectra. Additionally, 1ES\,1741+196 demonstrates hints of {\it extreme}-Syn nature in its XRT spectrum, while 1ES\,2322-409 appears as a standard HBL. The new sets of AstroSat and LAT data presented here have revealed more detailed spectral and variability properties of these sources.
  
We have conducted a detailed analysis of the SEDs of the selected sources, using both the contemporaneous data obtained from \astrosat\ and \fermi, as well as archived data available in various energy bands (Sec.~\ref{sec:observation}). We analyse the variability of the sources (Sec.~\ref{sec:analysis}) and model the various SEDs using different physical scenarios (Sec.~\ref{sec:spectralmod}). Firstly, we utilize the one-zone synchrotron self Compton (SSC) model developed by \cite{Boettcher13}, which has been successfully applied to a number of HBL sources. Secondly, we employ the electron-proton co-acceleration model developed by \cite{Zech2021} for certain {\it extreme}-TeV blazars. Lastly, we used the lepto-hadronic code \onehale\ \citep{Zacharias21,zacharias+22} that provides two different gamma-ray emission solutions: one lepto-hadronic case dominated by emission from secondary pairs, and another purely hadronic case with \g-ray emission dominated by proton synchrotron. Further information regarding the model descriptions can be found in Appendix~\ref{sec:codes}. We conclude in Sec.~\ref{sec:summary}.



\section{Observations and data analysis}\label{sec:observation}

We selected five HBL sources, 1ES\,0120+340, RGB\,J0710+591, 1ES\,1101-232, 1ES\,1741+196 and 1ES\,2322-409 for this work based on the available AstroSat data from our proposed observations. Four of them (all except 1ES\,2322-409) are known to exhibit the nature of eHBLs. 
We analyse AstroSat and the contemporaneous Fermi-LAT data centred at the AstroSat observation periods. The Fermi-LAT data are averaged for 4-6 years to attain a good fit statistic. The observation details are provided in Table \ref{table1} and the data analysis procedure are described in the following sections.

\subsection{AstroSat data: SXT, LAXPC and UVIT}\label{sec:astrosat}

AstroSat is a multiwavelength (MWL) space-based observatory that carries ﬁve scientiﬁc instruments on-board covering a wide range of energies from UV to hard X-rays. The 
instruments used in this work are: the Large Area X-ray Proportional Counters (LAXPC), the Soft X-ray focusing Telescope (SXT) and the Ultraviolet Imaging Telescope (UVIT). SXT is a focusing telescope capable of X-ray imaging and spectroscopy in the energy range 0.3 -- 8.0 keV 
\citep{2014SPIE.9144E..1SS,2016SPIE.9905E..1ES,2017JApA...38...29S}. The LAXPC instrument consists of three proportional counter units (LAXPC10, LAXPC20 and LAXPC30), 
providing coverage in the 3 -- 80~keV hard X-ray band \citep{2016SPIE.9905E..1DY,2017ApJS..231...10A}.
The UVIT onboard AstroSat is primarily an imaging telescope consisting of 3 channels in the visible and UV bands: FUV (130 ---180 nm), NUV (200 ---300 nm) and VIS (320 -- 550 nm) \citep{10.1117/12.924507, 2017AJ....154..128T}.
AstroSat observations of our selected sources were made as part of AO proposals, and both Level-1 and Level-2 data for each instrument are publicly available at the ISRO Science Data Archive\footnote{\href{https://astrobrowse.issdc.gov.in/astro$\_$archive/archive/Home.jsp}{astrobrowse.issdc.gov.in/astro$\_$archive/archive/Home.jsp}}. For this work we analyzed orbit-wise Level-1 science data for each of the instruments. 

\emph{\bf SXT:} The available SXT data were obtained in 
photon counting mode. The data from individual orbits were first processed with {\tt sxtpipeline} available in the SXT software (\emph{AS1SXTLevel2}, version 1.4b) package and merged into a single cleaned event file using the {\tt SXTEVTMERGER} tool. The analysis software and tools are available at the SXT POC website \footnote{\href{www.tifr.res.in/~astrosat$\_$sxt}{www.tifr.res.in/~astrosat$\_$sxt}}. 
The {\tt XSELECT} (V2.4d) package built-in to \emph{HEAsoft} was used to extract the source spectra in the energy range 0.3 -- 7~keV from the processed Level-2 cleaned event files.\\
As estimated by the {\tt sxtEEFmake} tool, a circular region of 16 arcmin radius centered on the source position that encompasses more than 95$\%$ of the source pixels was considered to generate spectral products. 
A standard background spectrum ({\tt "SkyBkg$\_$comb$\_$EL3p5$\_$Cl$\_$Rd16p0$\_$v01.pha"}) extracted from a composite product using a deep blank sky observation was used as a background (to avoid problems with the large Point Spread Function of SXT). A standard ancillary response file (ARF) of the individual sources were generated using the {\tt sxtmkarf} tool.
Further, we used a standard response file {\tt "sxt$\_$pc$\_$mat$\_$g0to12.rmf"} as RMF available at the SXT POC website. The extracted source spectra were then grouped using the {\tt grppha} tool to ensure  a minimum of 60 counts per bin. \\ 

\emph{\bf LAXPC:} The \emph{laxpcsoft} package available at the \emph{AstroSat} Science Support Cell (ASSC) website\footnote{\href{http://astrosat-ssc.iucaa.in/?q=sxtData}{astrosat-ssc.iucaa.in/?q=sxtData}} was used to process the Level-1 data. The standard data reduction steps were followed to generate the event files, standard GTI files of good time intervals to avoid Earth occultation and the South Atlantic Anomaly, and finally to extract the source spectra. To generate event and GTI files, we used the {\tt laxpc\_make\_event} and {\tt laxpc\_make\_stdgti} modules which are in-built in the \emph {laxpcsoft} package. Data from source free sky regions observed within a few days of the source observation were used to generate and model the background by using an appropriate scaling factor. Finally, the source spectra were generated by using the {\tt laxpc$\_$make$\_$spectra} tool. In case of faint sources like AGNs, the background estimation is not straightforward as the background starts dominating over the source counts. Therefore, the background was estimated from the 50 to 80 keV energy range where the background seems relatively steady. The data from the top layers of each LAXPC 10 and 20 units were reduced separately. LAXPC-30 data were discarded as recommended by the instrument team due to the continuous gain shift. However, the data from only LAXPC-20 in the energy range 3 -- 15~keV were used for the spectral analysis. \\

\emph{\bf UVIT:} We analysed UVIT data only for the sources 1ES\,1741+196 and RGB\,J0710+591. The data were available for 5 filters (3 NUV (NUVB13, NUVB4 and NUVN2) and 2 FUV (BaF2 and Silica)) for RGB\,J0710+591 and only for two FUV filters (BaF2 and Silica) for 1ES\,1741+196. The Level-1 data were processed with the \emph{UVIT Level-2 Pipeline} (Version 5.6), software accessible at the ASSC website and the standard data reduction procedures were followed. The pipeline generates the full-frame astrometry fits images which are corrected for flat-fielding and drift due to rotation. The fits images of individual orbits were then merged into a single fits image. To extract the counts from the fits images, aperture photometry was performed using the \emph{IRAF} (Image Reduction and Analysis Facility) software tool. An aperture of 50 pixels radius size was selected to do photometry which encompasses $\sim$ 98$\%$ of the source pixels. The extracted counts were then converted into fluxes for each filter using the unit conversion as suggested by \citep{2017AJ....154..128T}. The fluxes were then corrected for Galactic interstellar extinction \citep{1999PASP..111...63F} with their respective E$_{(B-V)}$ values taken from NED\footnote{\href{https://irsa.ipac.caltech.edu/applications/DUST/}{irsa.ipac.caltech.edu/applications/DUST/}}.

\subsection{ {\it Fermi}-LAT data}\label{sec:lat}

The Fermi-LAT data of the individual sources were taken from an epoch around the AstroSat observations, as listed in Table \ref{table1}. The Pass 8 (P8R3) data were downloaded from the LAT data center\footnote{\href{https://fermi.gsfc.nasa.gov/cgi-bin/ssc/LAT/LATDataQuery.cgi}{fermi.gsfc.nasa.gov/cgi-bin/ssc/LAT/LATDataQuery.cgi}} with a 15 degree search radius. The publicly available software {\tt Fermitools} (version 2.0.8) and the python package {\tt fermipy} (version v1.0.1) \footnote{\href{http://fermipy-readthedocs.io/en/latest/}{fermipy-readthedocs.io/en/latest/}}\citep{2017ICRC...35..824W} were used to perform the data analysis. We used event class 128 (P8R3 SOURCE), event type 3 (i.e. FRONT+BACK), a standard data quality selection criteria ($DATA\_QUAL>0\,\&\&\,LAT\_CONFIG==1$) and the energy range 0.3 -- 300~GeV for all datasets. The zenith angle was set at the maximum value of 90$^\circ$ to avoid contamination due to Earth's limb and spacecraft events. The P8R3\_SOURCE\_V2 instrument response functions (IRFs), the diffuse Galactic interstellar emission (gll\_iem\_v07.fits) and the isotropic emission (iso\_P8R3\_SOURCE\_V2\_v1.txt) were used and we included all sources listed in the fourth Fermi-LAT catalog (4FGL) \citep{Abdollahi+2020} in the background model. While performing the spectral fit the parameters of all sources within 3$^\circ$ of the source were set free. The SEDs were then generated using the best-fit model parameters using the standard sed method in fermipy with 2 bins per decade.

\subsection{Archival MWL data}\label{sec:mwl}

Archival multiwavelength (MWL) data in the optical -- UV and TeV were taken from \cite{Costamante+2018} for the sources RGBJ\,0710+591 (WISE and VERITAS data) and 1ES\,1101-232 (WISE and H.E.S.S. data); \cite{Ahnen+2017}, \cite{Abeysekara+2016} and \cite{Wierzcholska+2020} for 1ES\,1741+196 (MAGIC, VERITAS and WISE+2MASS+ATOM data); and \cite{HESS+2322} for 1ES\,2322-409 (WISE and H.E.S.S.). TeV spectra were corrected for EBL absorption, except for 1ES\,2322-409. We further analysed the quasi-simultaneous optical -- UV and X-ray data from Swift UVOT and XRT and NuSTAR for comparison. These data were analysed using standard data analysis procedures and the pipelines {\tt uvotsource}\footnote{\href{https://www.swift.ac.uk/analysis/uvot/}{www.swift.ac.uk/analysis/uvot/}}, {\tt xrtpipeline}\footnote{\href{https://www.swift.ac.uk/analysis/xrt/}{www.swift.ac.uk/analysis/xrt/}}, 
{\tt nupipeline}\footnote{\href{https://heasarc.gsfc.nasa.gov/docs/nustar/analysis/}{heasarc.gsfc.nasa.gov/docs/nustar/analysis/}}.  




\begin{table*}
\centering
	\caption{Details of the observations from various instruments of \emph{AstroSat} missions.} 
\begin{tabular}{ccccccc}
\hline \hline 
 
	Source & Observation ID & Observation date & Exposure (SXT) & Instruments (/filters) & State \\ 
	& &    (yyyy-mm-dd) & (ks)&  & \\

\hline \hline
	\textbf{1ES\,0120+340}& A05-185T04-9000002548 & 2018-12-01 & 112.12 & SXT, LAXPC-20 & A1 \\ 
 & A05-185T04-9000002552 & 2018-12-06 & 20.68 & &A2 \\
\hline 
	\textbf{RGB\,J0710+591}& A02-085T02-9000000808 & 2016-11-19 & 20.25 & UVIT (2 FUV), SXT, LAXPC-20 & A1 \\ 
\hline 

	\textbf{1ES\,1101-232}& G06-086T02-9000000936 & 2016-12-28 & 70.78 & SXT, LAXPC-20 & A1 \\ 
\hline 
	& A05-163T01-9000002820 & 2019-03-27& 33.94 & & A1  \\
	\textbf{1ES\,1741+196}&A05-163T01-9000003010 & 2019-07-03& 33.12 & SXT, LAXPC-20 & A2 \\ 
	&A05-163T01-9000003118 & 2019-08-22 & 24.8 & UVIT (3 NUV, 2 FUV) & A3  \\
\hline

	\textbf{1ES\,2322-409}& A09-147T01-9000003754 & 2020-07-03 & 44.21 & SXT, LAXPC-20 & A1 \\ 
\hline \hline\\
\end{tabular}
\label{table1}
\end{table*}



\section{Spectral and temporal variability}\label{sec:analysis}

We analysed simultaneous SXT and LAXPC spectra from one single pointing observation each for RGB\,J0710+591, 1ES\,1101-232 and 1ES\,2322-409, 2 pointings for 1ES\,0120+340, and 3 pointings for 1ES\,1741+196 (as shown in Table \ref{table1}). A1, A2 and A3 denote different X-ray spectral states for the sources 1ES\,0120+340 and 1ES\,1741+196, respectively, used for spectral analysis. We fitted these combined spectra with single power-law and log-parabola spectral models. 
The spectral fittings were performed for each observation separately using the {\tt XSPEC} (Version 12.9.1) software package \citep{Arnaud+1996} distributed with \emph{HEASoft}. In addition, we used the ISM absorption model \emph{TBabs} available in {\tt XSPEC} with the appropriate choice of the neutral hydrogen column density (N$_H$) value fixed throughout the analysis. The N$_H$ values were estimated with an online tool\footnote{\href{https://heasarc.gsfc.nasa.gov/cgi-bin/Tools/w3nh/w3nh.pl}{heasarc.gsfc.nasa.gov/cgi-bin/Tools/w3nh/w3nh.pl}} using the LAB survey map \citep{Kalberla+2005}. A best-fit nominal gain offset of 0.3 keV as determined using the {\tt gain fit} option with a fixed gain slope of 1 was used, as recommended by the SXT instrument team. This significantly improves the fit statistics. Once the best-fit gain parameters were decided, we fixed these throughout the spectral fitting to save computation time while calculating the error bars. An additional systematic error of 3$\%$ was used for joint SXT-LAXPC spectral fits to minimize background uncertainties as recommended. To account for the relative cross-calibration uncertainties between the two X-ray instruments, a multiplicative constant factor was added to the spectral model and was kept fixed for SXT and free to vary for the LAXPC instrument.

\begin{figure*}
\includegraphics[width=0.45\textwidth]{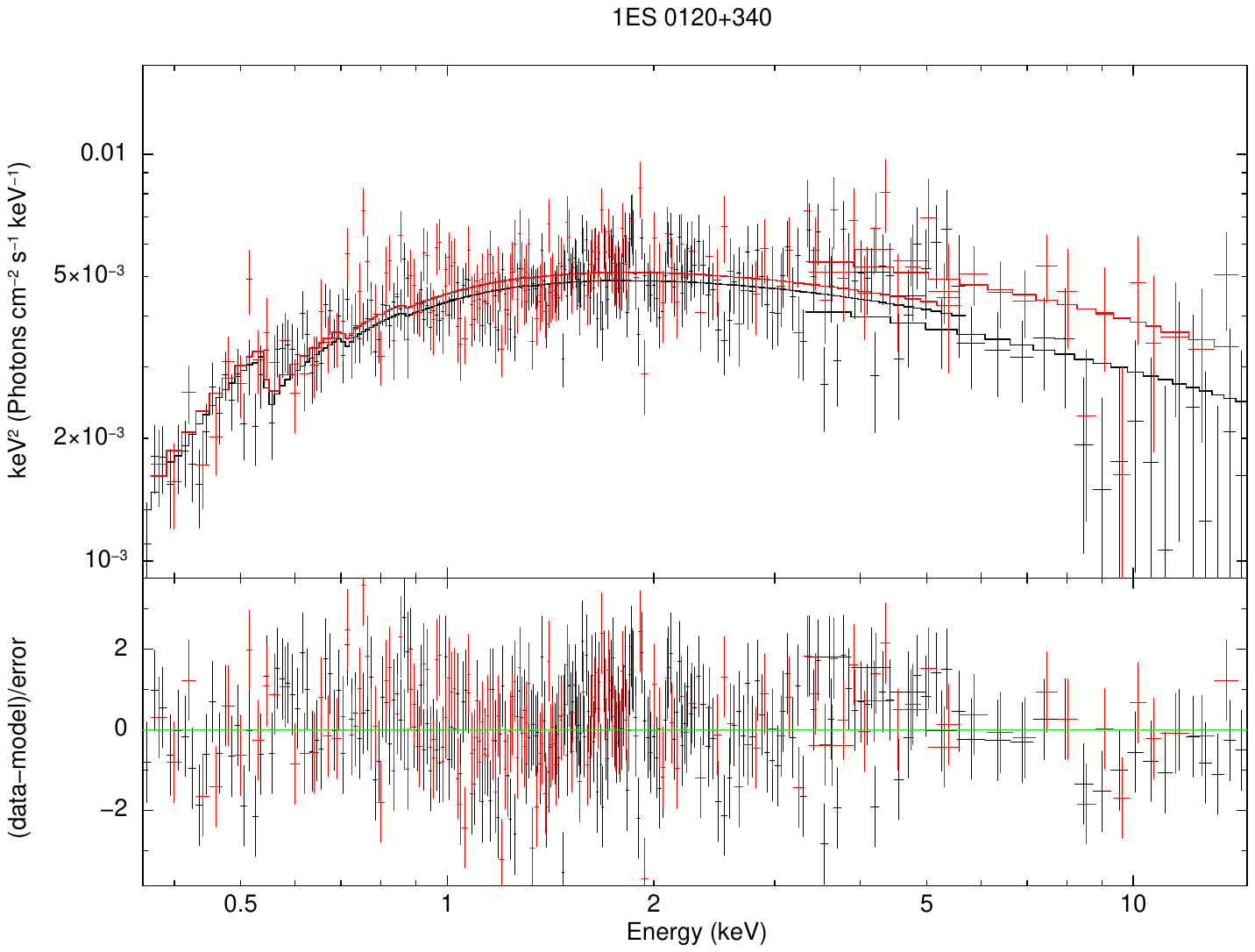}
\includegraphics[width=0.45\textwidth]{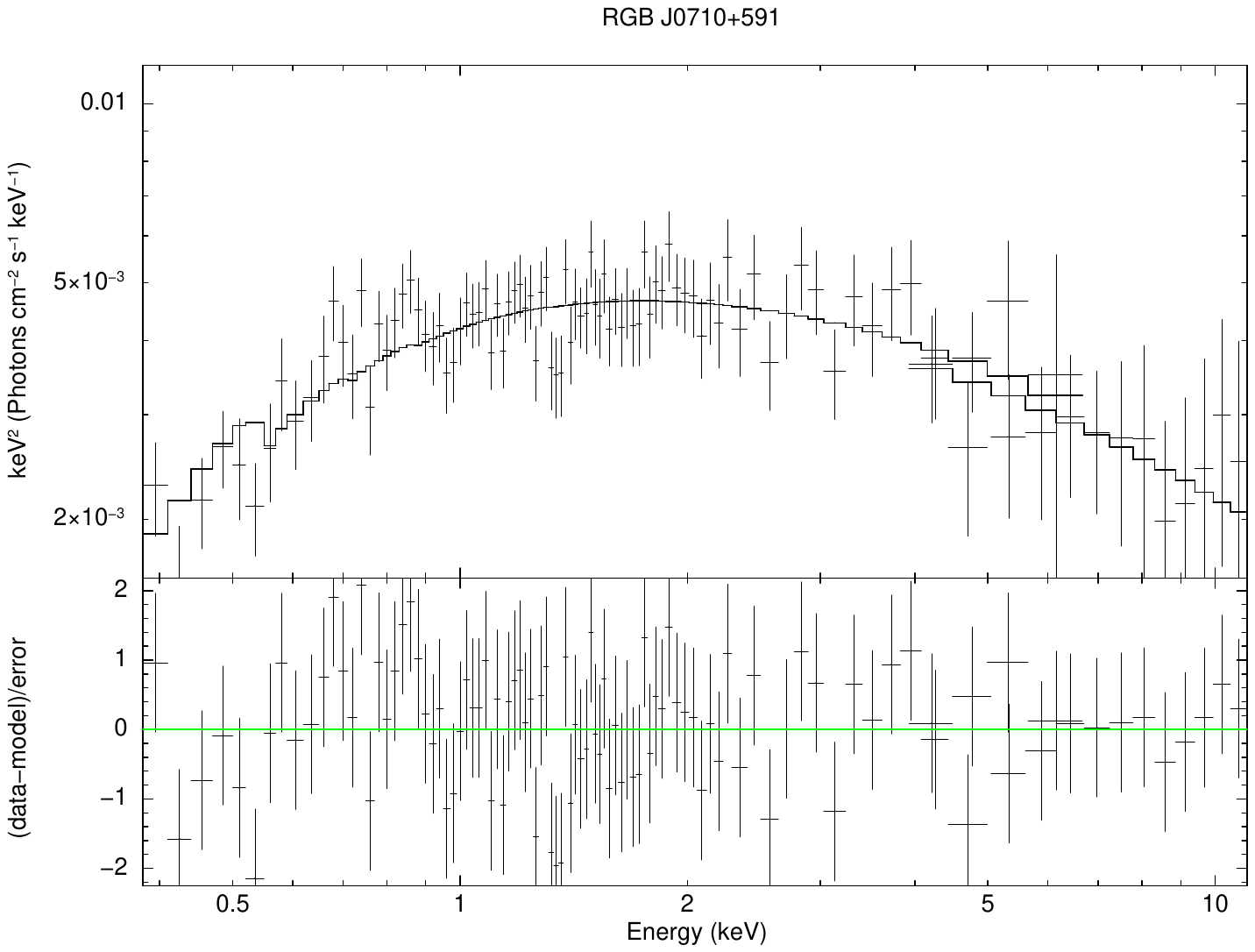}\\
\includegraphics[width=0.45\textwidth]{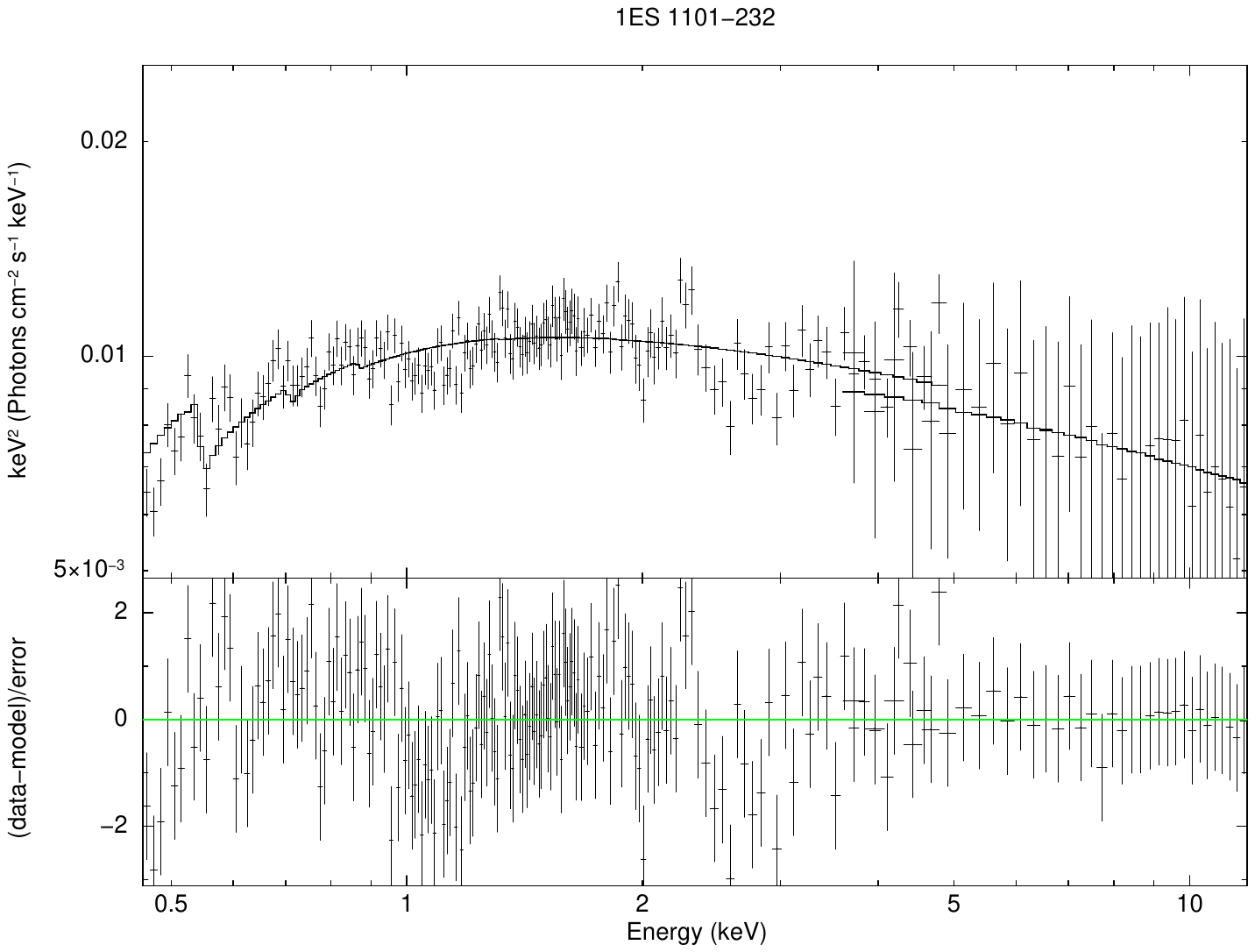}
\includegraphics[width=0.45\textwidth]{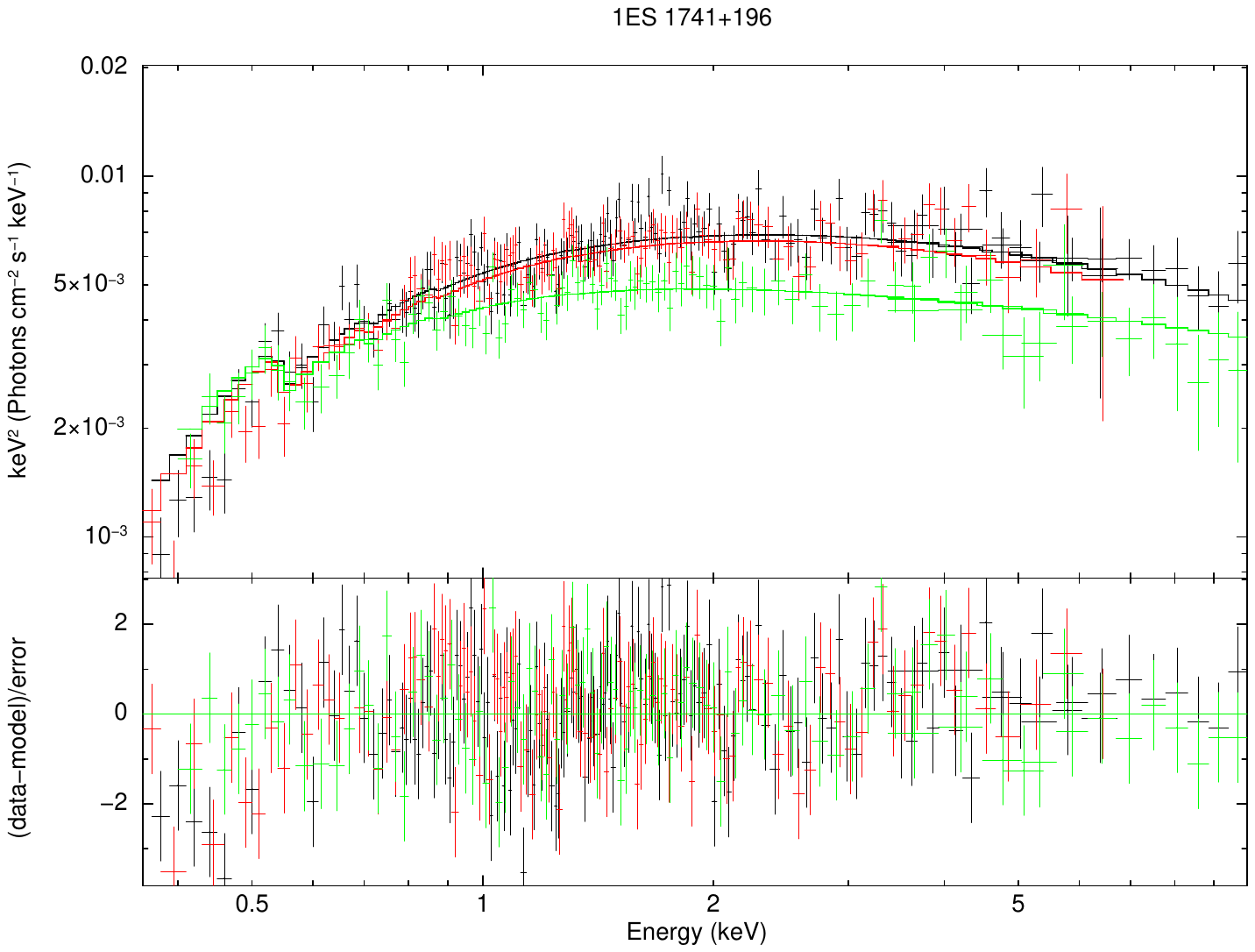}\\
\includegraphics[width=0.45\textwidth]{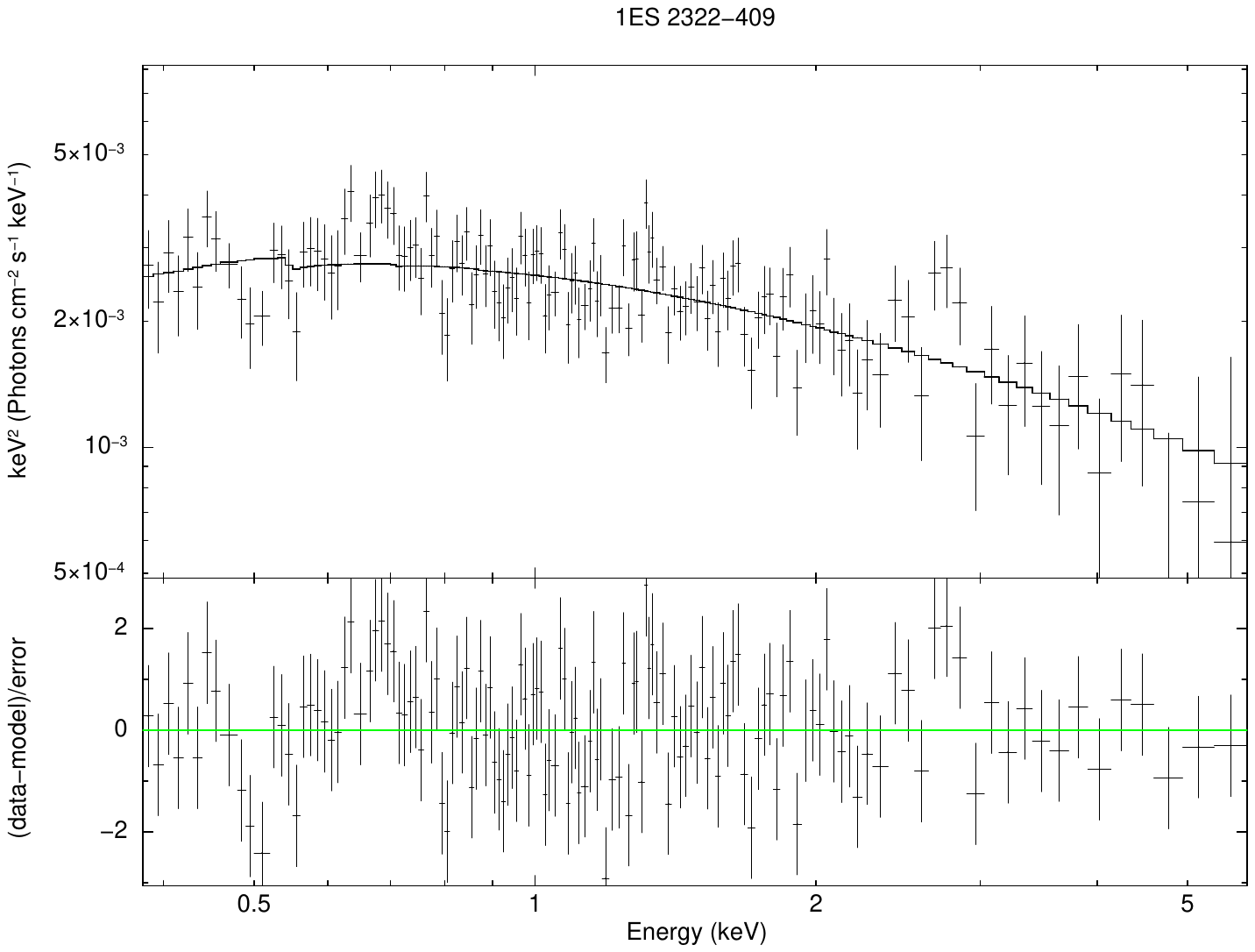}
\caption{\label{fig:xray_fit} Combined SXT and LAXPC-20 spectrum of 1ES\,0120+340 , RGB\,J0710+591 (A1: black; A2:red), 1ES\,1101-232, 1ES\,1741+196 (A1: black, A2: red; A3: green), and 1ES\,2322-409(SXT only), [from left to right in top and bottom], fitted using the \emph{TBabs*logparabola} model.
}  
\end{figure*}

Initially, we attempted to fit the X-ray spectra of all sources using a simple power-law (PL) model, but this yielded poor fits, indicating the presence of intrinsic curvature in the spectra. Indeed, a log-parabola (\emph{logpar} model in {\tt XSPEC}) provides good statistical fits. The best-fit \emph{logpar} model parameters along with their uncertainties estimated within a 90\%
confidence range are reported in Table \ref{table2}, and the corresponding spectral fits are shown in Figure~\ref{fig:xray_fit}. In all cases, the combined SXT and LAXPC spectra up to 10 keV are able to pin down the location of the synchrotron peaks well within the observed energy range. The peak energies ($\epsilon_p$) are estimated using the \emph{eplogpar} model included in {\tt XSPEC}. For the first four sources (ref. to Table \ref{table2}), we observe hard spectral indices ($\alpha <2$) and synchrotron-peak energy values above $1\,$keV. 
Especially for 1ES\,0120+340 and 1ES\,1741+196, the \astrosat\ observations provide the first well-constrained estimation of the synchrotron peak values. However, 1ES\,2322-409 is an exception, satisfying the criteria of a regular HBL with a relatively soft spectral index and synchrotron peak located below $0.3\,$keV.

\begin{figure}
\includegraphics[width=0.5\textwidth]{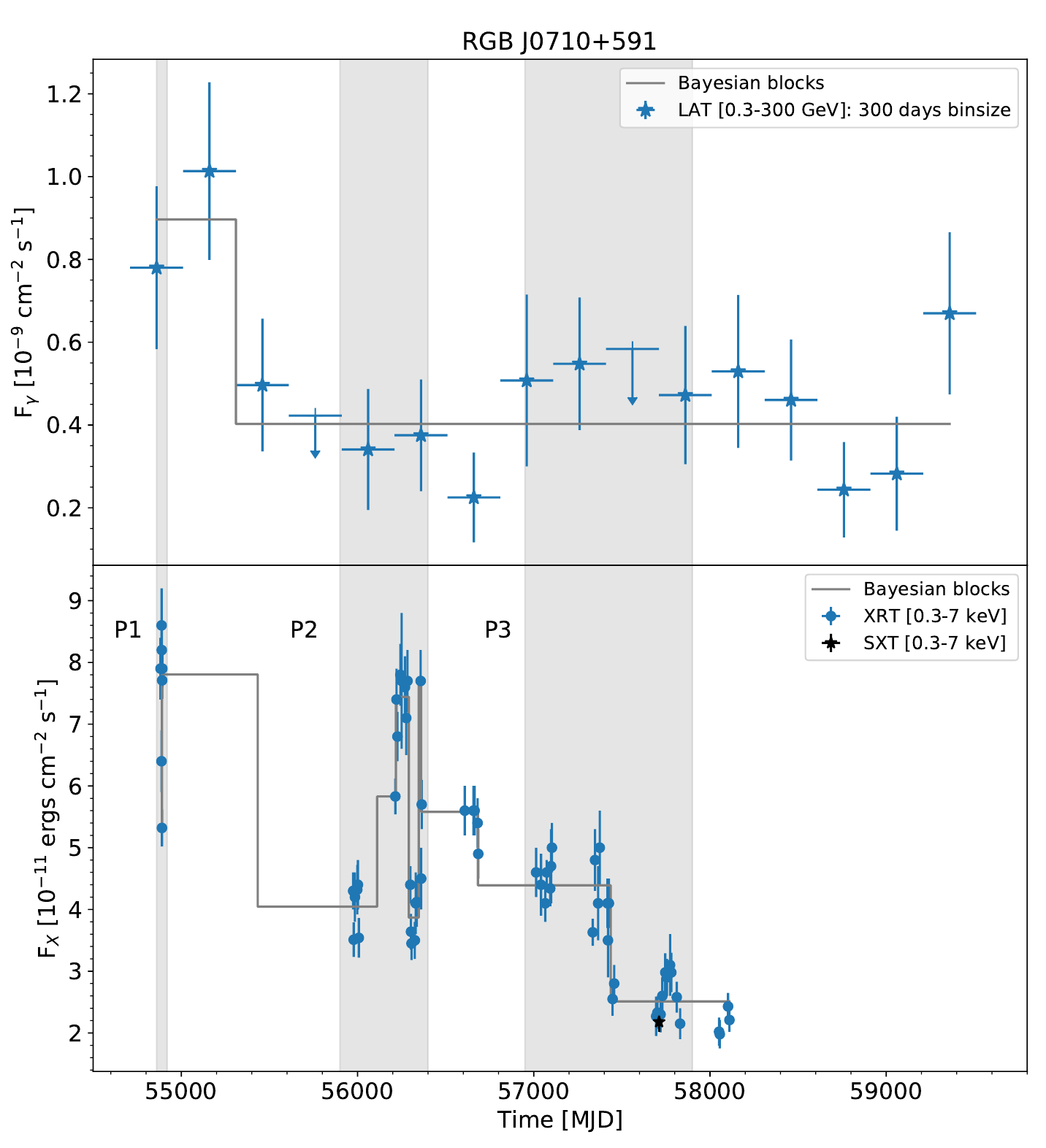}
\caption{Lightcurves of RGB\,J0710+591 overlaid with the Bayesian blocks representation. {\bf Top panel:} Fermi-LAT in the energy range 0.3-300 GeV from 2008/09-2021/09 in 300-day bins in units of 10$^{-9}$ cm$^{-2}$ s$^{-1}$. The upper limits correspond to 3$\sigma$ limits. The dotted line represents the constant-fit. 
{\bf Bottom Panel:} Swift-XRT (in blue) and AstroSat-SXT (in black) in the energy range 0.3-7 keV in units of 10$^{-11}$ erg cm$^{-2}$ s$^{-1}$.
}
\label{fig:j0710_lc}
\end{figure}

X-ray flux variability is seen in some of the sources over various time scales. For example, RGB\,J0710+591 shows a significant spectral transition with a strongly increasing spectral curvature ($\beta$ increased by a factor $\sim 1.8$) and a marginal change in its spectral index and flux over a period of one year \citep[cf.,][]{Costamante+2018, Goswami+2020}. The X-ray lightcurve in the energy range 0.3 -- 7~keV obtained for the period 2009/02 -- 2017/12 (MJD 54882.18 -- 58110.26) is shown in the bottom panel of Fig.~\ref{fig:j0710_lc}. We first apply the Bayesian blocks algorithm \citep{Scargle_2013} implemented in {\tt astropy} \footnote{\href{https://docs.astropy.org/en/stable/api/astropy.stats.bayesian\_blocks.html}{docs.astropy.org/en/stable/api/astropy.stats.bayesian\_blocks}} 
to the long lightcurve. We use the option of ‘measures’ in the fitness function and false alarm probability p0 = 0.01. We observe long-term flux variations:: high flux activity during 2009/02 -- 2009/03 (period P1) and 2012/02 -- 2013/01 (period P2), and low flux activity during 2015/01 -- 2017/12 (period P3). We further characterize the flux variations in these periods by their doubling/halving timescales ($\Delta t_D$/$\Delta t_H$) which are given as $\Delta t = t\,\times\,\ln{2}/|\ln{(F2/F1)}|$ \citep{2013ApJ...766L..11S}. Here, $F1$ and $F2$ are the fluxes observed at a time interval of duration $t$.
The highest flux is observed during period P1 with flux $(8.6\pm0.6) \times 10^{-11}$~erg~cm$^{-2}$~s$^{-1}$. The flux during P2 shows a rise followed by a sharp decline and peaks at a flux $(7.8\pm0.5) \times 10^{-11}$~erg~cm$^{-2}$~s$^{-1}$ with $\Delta t_D = 14.5\pm2.68$~days and $\Delta t_H = 0.92\pm0.11$~days. The period P3 shows a slowly decaying trend with an estimated halving timescale of $\Delta t_D = 29.7\pm5.31$~days. The overall variation can further be characterised by the fractional variability amplitude \citep[following the definition by][]{2003MNRAS.345.1271V}. The mean fractional variability in the long-term lightcurve is $\sim$ 39.6\%. 

For 1ES\,1741+196, we observe significant X-ray flux variations over a long observation period from 2010/07 -- 2020/10. The mean fractional variability in the long-term lightcurve is $\sim 32.2$~\%.

To obtain significant spectra in the \g-ray band detected by \fermi, our sources require integration times of several years. For every source, we obtained spectra focused on the \astrosat\ observations (see Table~\ref{table1}). In the case of RGB\,J0710+591, 1ES\,1101-232, and 1ES\,1741+196, we also computed spectra that coincide with the archival VHE and X-ray (XRT/NuSTAR) observations used for the SED analysis. The \fermi\ integration times and spectral fit parameters are provided in Table~\ref{table3}.
The spectra are fitted with a power-law in the energy range 0.3 -- 300~GeV. The absorption effects due to EBL at these energies are likely negligible for the observed sources and hence, no corrections have been made. The low photon indices ($\Gamma_{LAT} \leq 1.6$) of the observed spectra for 1ES\,0120+340, RGB\,J0710+591 and 1ES\,1101-232 indicate that the high-energy peak may be located at energies above $1\,$TeV, leading to their classification as {\it extreme}-TeV blazars. While the spectral indices in 1ES\,1741+196 and 1ES\,2322-409 are below $2.0$, the soft VHE spectra (coupled with their X-ray spectra) result in their classification as an {\it extreme}-Syn blazar and an HBL, respectively.
Overall, the individual \fermi\ spectra show consistent behaviour over extended periods of observations, except for RGB\,J0710+591. 
A hint of high flux activity is noticed in its extended lightcurve (upper panel of Figure~\ref{fig:j0710_lc}) coinciding with the period P1. This is also illustrated by the presentation of Bayesian block generated with a p0 value of 0.01.



\begin{table*}
\centering
\footnotesize
	\caption {Best-fit log-parabola model parameters of joint SXT-LAXPC spectrum using \emph{TBabs*logpar} model. col.[1]: Source name and abbreviation for different spectra, col.[2]: galactic N$_{H}$ value in the units of $10^{20}$ cm$^{-2}$, col.[3]: X-ray energy range in keV used for spectral fitting, col.[4]: The relative cross-normalization constant between SXT and LAXPC instruments. This parameter was fixed at 1.0 for SXT and kept free for LAXPC data while performing joint SXT-LAXPC spectral fit, col.[5]: spectral index is estimated at 1\,keV, col.[6]: curvature parameter, col.[7]: synchrotron peak energy in keV estimated using \emph{TBabs*eplogpar} model, col.[8]: 2-10 keV average flux (F) is in units of 10$^{-11}$ erg cm$^{-2}$ s$^{-1}$, col.[9]: reduced chi-square and degrees of freedom values. The errors are estimated within 90$\%$ confidence range based on the criterion used in {\tt XSPEC}.} 
\begin{tabular}{ccccccccc}
\hline \hline 

	Name & N$_{H}$ & Energy &  Constant &  $\alpha$ & $\beta$ & $\epsilon_p$ (keV) &  F$\rm\,_{2-10\,keV}$ & Stats.(chi sq./dof) \\

	 [1] & [2] &  [3] & [4] & [5] & [6] &  [7]  & [8] & [9] \\
\hline 
	{\bf 1ES\,0120+340} & 5.23& 0.3 -- 15  & & & & & & \\ 
	A1 &  && 1.05$^{+0.16}_{-0.15}$ & 1.93 $^{+0.04}_{-0.04}$  & 0.27 $^{+0.10}_{-0.09}$ & 1.47 $^{+0.20}_{-0.16}$  & 1.03 $^{+0.02}_{-0.04}$ & 1.19 (266.74/224)\\
	A2 &  && 1.06$^{+0.18}_{-0.15}$ & 1.89 $^{+0.06}_{-0.06}$  & 0.22 $^{+0.12}_{-0.12}$ & 1.69 $^{+0.20}_{-0.22}$  & 1.20 $^{+0.07}_{-0.07}$ & 1.16 (186.84/160)\\
\hline \\
	{\bf RGB\,J0710+591 } & 4.48& 0.3 -- 10  & & & & & & \\
	A1 &  && 0.90$^{+0.21}_{-0.17}$ & 1.77 $^{+0.07}_{-0.07}$  & 0.52 $^{+0.16}_{-0.17}$ & 1.46 $^{+0.26}_{-0.19}$  &  0.95 $^{+0.05}_{-0.04}$ & 0.79 (76.59/97)\\
\hline \\
	{\bf 1ES\,1101-232} &5.51& 0.3 -- 15  & & & & & & \\
	A1 &  && 1.04$^{+0.14}_{-0.13}$ & 2.09 $^{+0.02}_{-0.02}$  & 0.17 $^{+0.05}_{-0.05}$ & 0.85 $^{+0.27}_{-0.20}$  &  2.14 $^{+0.06}_{-0.05}$ & 1.15 (420.90/366)\\
	\hline \\
	{\bf 1ES\,1741+196} & 7.32& 0.3 -- 10 & & & & & & \\
	A1 &  && 1.08 $^{+0.13}_{-0.13}$ & 1.75 $^{+0.05}_{-0.05}$  & 0.41 $^{+0.10}_{-0.09}$ & 2.07 $^{+0.22}_{-0.20}$  &  1.54 $^{+0.07}_{-0.03}$  & 1.18 (281.50/238)\\
	A2 &  && 1.17 $^{+0.54}_{-0.48}$  &1.73 $^{+0.06}_{-0.05}$ & 0.45 $^{+0.13}_{-0.12}$  & 1.98 $^{+0.21}_{-0.19}$  &  1.45 $^{+0.14}_{-0.05}$ & 1.09 (240.58/219) \\
	A3 &  && 0.91 $^{+0.17}_{-0.14}$ & 2.01 $^{+0.06}_{-0.06}$ & 0.16 $^{+0.14}_{-0.13}$& 0.98 $^{+0.27}_{-0.38}$  &  1.10 $^{+0.05}_{-0.06}$ & 0.87 (157.93/181)\\
\hline \\

	{\bf 1ES\,2322-409} &1.57 & 0.3 -- 6  & & & & & & \\
	A1 &  && -- & 2.34 $^{+0.05}_{-0.05}$  & 0.37 $^{+0.20}_{-0.19}$ & 0.33 $^{+0.27}_{-0.20}$  &  0.32 $^{+0.07}_{-0.04}$ & 1.17 (161.05/137)\\
	
\hline \hline \\
\end{tabular} \\
\label{table2}
\end{table*}

\begin{table}
\centering
\footnotesize
	\caption {Best-fit power-law model parameters of observed Fermi-LAT spectrum. col.[1]: source name with LAT integration time and abbreviation for different spectra, col.[2]: test-statistic values, col.[3]: photon index $1\sigma$ error, col.[4]: integrated flux in the energy range 0.3-300 GeV with $1\sigma$ error in units of $10^{-9}$ cm$^{-2}$ s$^{-1}$}.
\begin{tabular}{lccc}
\hline \hline 
Name & TS & PL Index & Flux \\

[1]&[2]&[3] &[4] \\
\hline 
{\bf 1ES\,0120+340} & && \\
2015/01-2021/01 \ [L1] & 53.32 & 1.35$\pm$0.25 & 0.11$\pm$0.04\\
{\bf RGB\,J0710+591} &  & & \\
2008/09-2009/08 \ [L1] & 92.99 & 1.45$\pm$0.14 & 2.06$\pm$0.83 \\
2013/01-2016/12 \ [L2] & 102.20 & 1.67$\pm$0.11 & 0.32$\pm$0.06 \\
2017/01-2021/01 \ [L3] & 139.87 & 1.76$\pm$0.12 & 0.54$\pm$0.07 \\
{\bf 1ES\,1101-232} & && \\
2009/01-2015/01 \ [L1] & 52.05 & 1.63$\pm$0.19 & 0.31$\pm$0.09 \\
2015/01-2019/01 \ [L2] & 63.3 & 1.65$\pm$0.15 & 0.35$\pm$0.08 \\
{\bf 1ES\,1741+196} && & \\
2010/01-2011/01 \ [L1] & 45.72 & 1.84$\pm$0.15 & 5.45$\pm$2.32 \\
2017/01-2021/01 \ [L2] & 161.25 & 1.86$\pm$0.08 & 4.97$\pm$1.21 \\
{\bf 1ES\,2322-409} & & & \\
2019/01-2021/01 \ [L1] & 787 & 1.79$\pm$0.05 & 7.17$\pm$0.97 \\
\hline\hline
\end{tabular}
\label{table3}
\end{table}


\begin{figure*}
\includegraphics[width=0.45\textwidth]{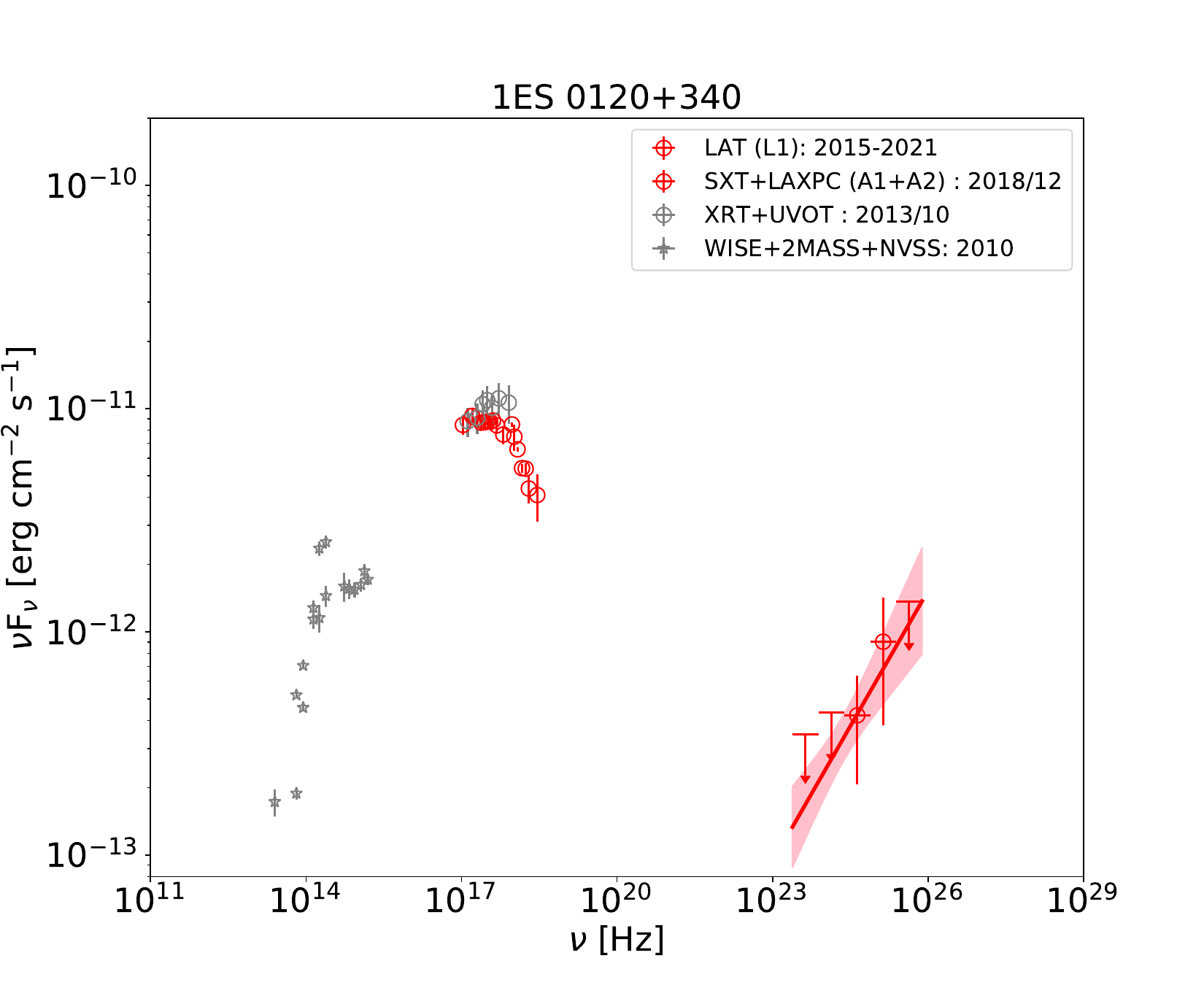}
\includegraphics[width=0.45\textwidth]{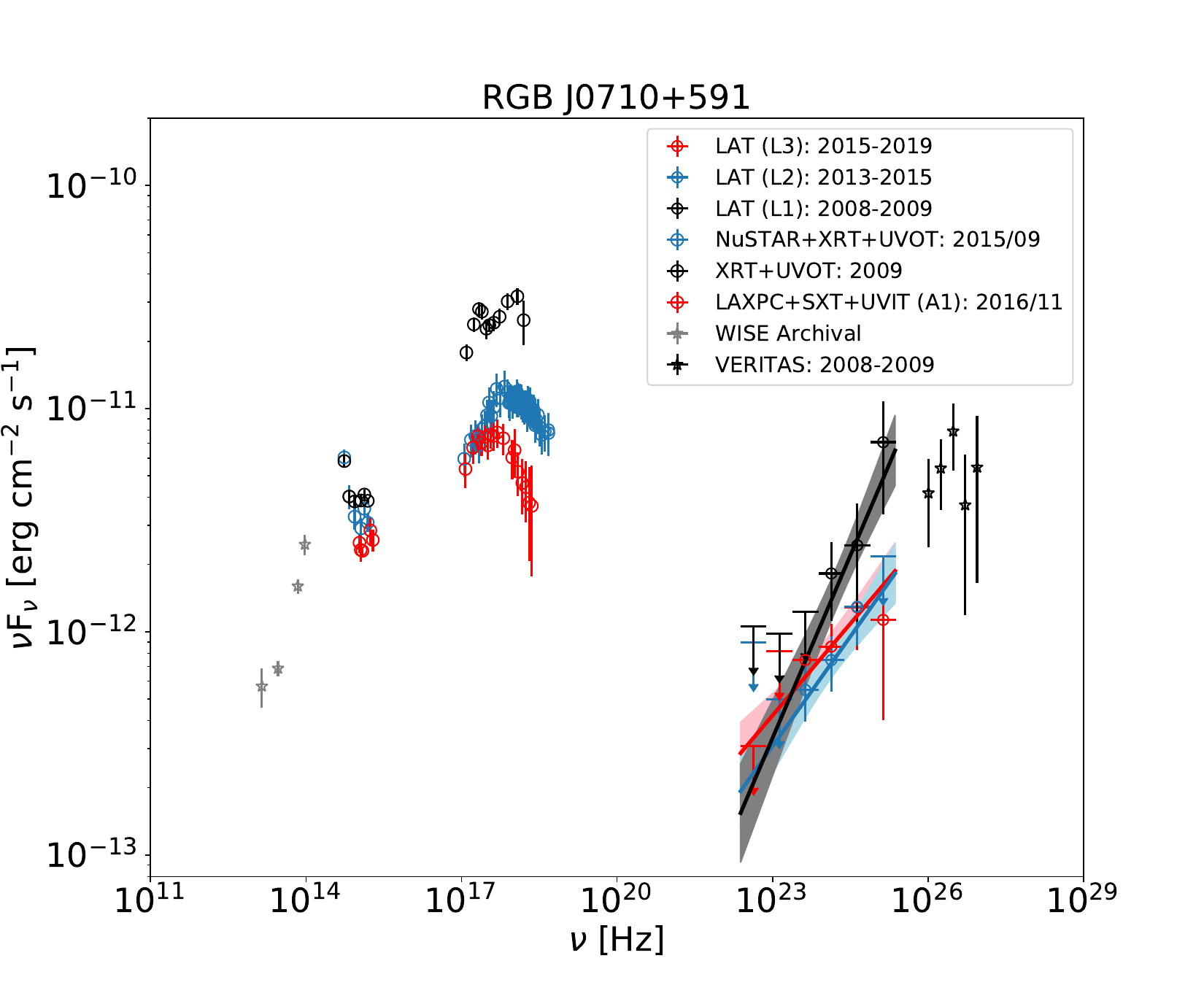}\\
\includegraphics[width=0.45\textwidth]{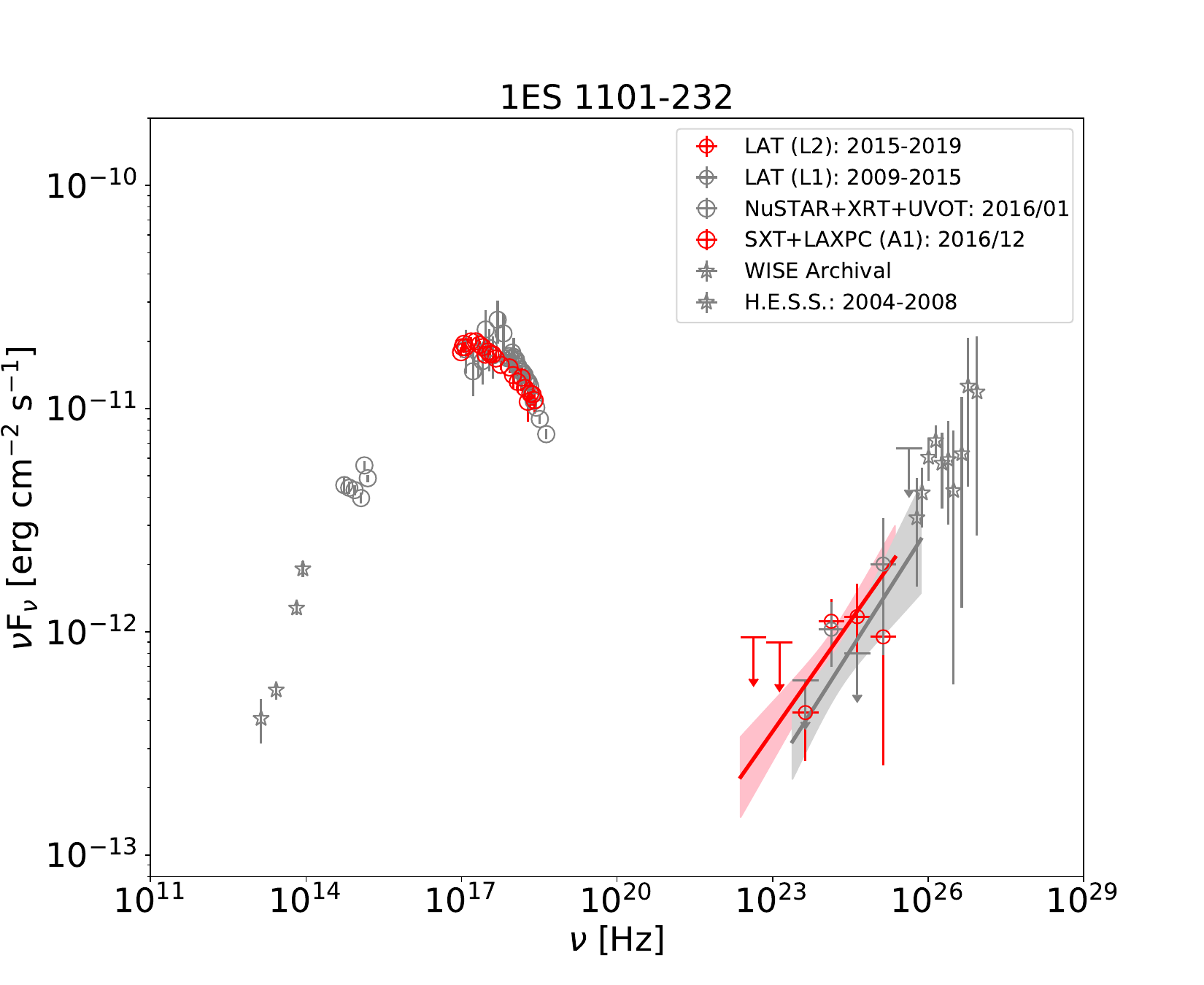}
\includegraphics[width=0.45\textwidth]{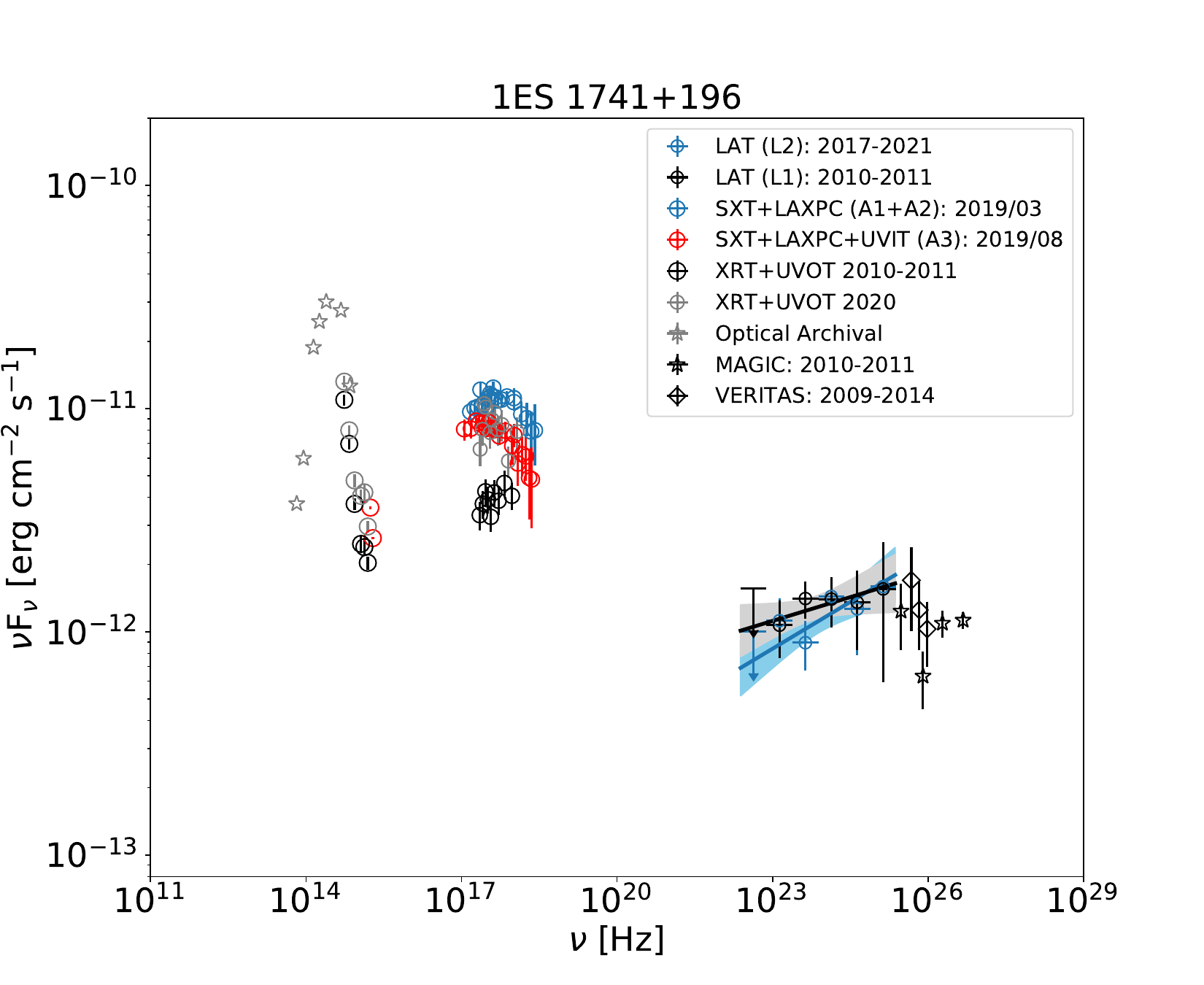}\\
\includegraphics[width=0.45\textwidth]{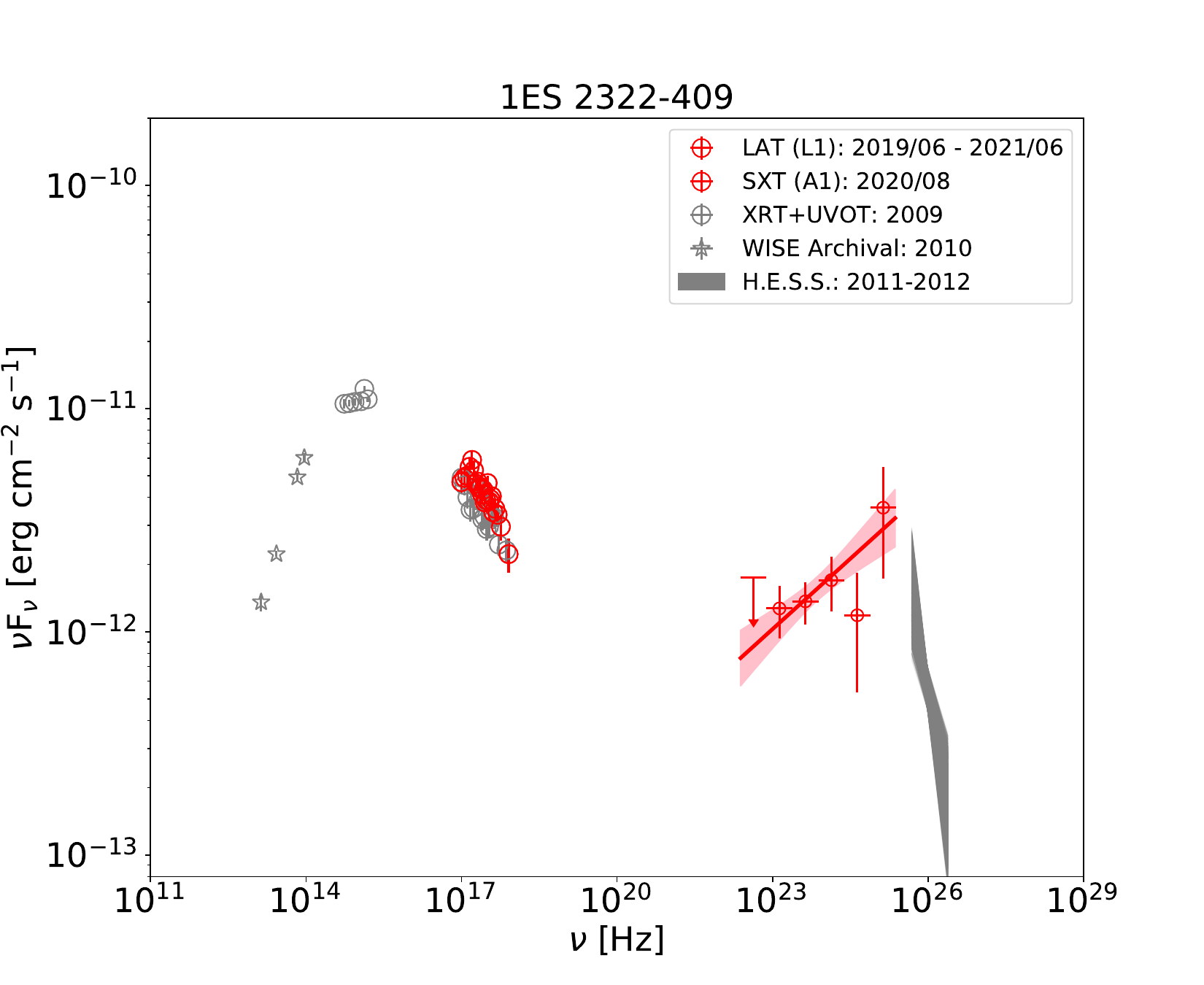}
\caption{\label{fig:sseddt} The MWL SEDs of 1ES\,0120+340,  RGB\,J0710+591, 1ES\,1101-232, 1ES\,1741+196, and 1ES\,2322-409, respectively, with details given in the legends. The colored spectra follow the definition given in Tab.~\ref{tab:modelsources}.
}
\end{figure*}




\section{Spectral modeling} \label{sec:spectralmod}

The previous section has shown that some sources are variable (RGB\,J0710+591 and 1ES\,1741+196), while the others show seemingly constant flux. In order to properly model the SEDs of the five sources, 
we will first define the data sets to be modeled in Sec.~\ref{sec:moddata}, followed by a derivation of constraints in Sec.~\ref{sec:modconstraints}. We then proceed with modeling the data sets with four different set-ups (Sec.~\ref{sec:modeling}).

\subsection{Data sets} \label{sec:moddata}


\begin{table*}
\centering
\footnotesize
	\caption {Source power-law indices of the SED in the optical/UV-to-X-ray part, $\alpha_{ox}$, and the HE \g-ray part, $\alpha_{\g}$, as well as the time ranges and collected data sets for the modeling. The definitions of the \astrosat\ and \fermi\ spectra are in Tabs.~\ref{table2} and~\ref{table3}, respectively, while the additional archival data sets can be found in Fig.~\ref{fig:sseddt}.} 
\begin{tabular}{lccccl}
\hline \hline 
	Source & Spectrum & $\alpha_{ox}$ & $\alpha_{\g}$ & Time range & Main data   \\
\hline 
1ES\,0120+340  & S1 & $0.38\pm0.06$ & $0.65\pm0.25$ & 2015/01-2021/01 & A1+A2,L1 \\
\hline
RGB\,J0710+591 & S1 & $0.40\pm0.07$ & $0.55\pm0.14$ & 2008/09-2009/08 & UVOT09,XRT09,L1,VERITAS09 \\
               & S2 & $0.15\pm0.10$ & $0.33\pm0.11$ & 2013/01-2016/12 & UVOT13,XRT13,NuSTAR13,L2 \\
               & S3 & $0.23\pm0.09$ & $0.24\pm0.12$ & 2017/01-2021/01 & UVIT(A1),A1,L3 \\
\hline
1ES\,1101-232  & S1 & $0.31\pm0.04$ & $0.31\pm0.19$ & 2015/01-2019/01 & A1,L2 \\
\hline
1ES\,1741+196  & S1 & $0.04\pm0.05$ & $0.16\pm0.15$ & 2010/01-2011/01 & UVOT10,XRT10,L1,MAGIC10 \\
               & S2 & $0.26\pm0.05$ & $0.14\pm0.08$ & 2017/01-2021/01 & A1+A2,L2 \\
               & S3 & $0.23\pm0.07$ & $0.14\pm0.08$ & 2017/01-2021/01 & UVIT(A3),A3,L2 \\
\hline
1ES\,2322-409  & S1 &$-0.21\pm0.07$ & $0.21\pm0.05$ & 2019/01-2021/01 & A1,L1 \\
\hline\hline
\end{tabular} \\
Notes: The SED index $\alpha$ is defined as $\nu F_{\nu}\propto \nu^{\alpha}$ and related to the photon index $\Gamma_p$ by $\alpha = 2-\Gamma_p$. \\
\label{tab:modelsources}
\end{table*}

We compiled data sets which are as complete and as contemporaneous as possible. The data sets are defined in Tab.~\ref{tab:modelsources} and the corresponding SEDs are shown in Fig.~\ref{fig:sseddt}. As many bands are not covered simultaneously with any of the defined data sets, we also gathered additional non-contemporaneous data, which we label archival data and show in gray in Fig.~\ref{fig:sseddt}.

In 1ES\,0120+340, the two \astrosat\ spectra are compatible with each other, while they seem to be slightly softer than the earlier \swift-XRT spectrum. However, the LAXPC spectrum indicates that the cut-off is beyond $10^{17}\,$Hz. We will thus consider only one spectrum for the modeling. Unfortunately, no VHE \g-ray data are publicly available. Hence, the high energy peak of this \eTeV\ source is not constrained well.

RGB\,J0710+591 is a bona-fide \eTeV\ source that is now firmly established to be variable in both spectral components. We collect three MWL spectra for modeling: Spectrum 1 (black in Fig.~\ref{fig:sseddt}) comprising of data from \swift, \fermi\ and VERITAS; Spectrum 2 (blue in Fig.~\ref{fig:sseddt}) containing data from \swift, \nustar\ and \fermi; and Spectrum 3 (red in Fig.~\ref{fig:sseddt}) with data from \astrosat\ and \fermi. In the X-ray domain, Spectrum 1 is both higher in flux and harder than Spectra 2 and 3. Additionally, from Spectrum 2 to 3 the peak frequency reduces.
As the optical/UV fluxes seemingly do not change --- possibly due to a significant contribution by the host \citep{veritas10} --- the spectral index describing the spectrum from the UV to the X-ray domain drops from Spectrum 1 to Spectra 2 and 3.
The \fermi\ spectra also indicate spectral variability with a softening throughout. The connection to the VHE \g-ray spectrum is not perfect for either \fermi\ spectrum, suggesting that the VHE \g-ray spectrum is also variable. 

In the \eTeV\ source 1ES\,1101-232, the X-ray spectrum seems stable between the different observations from \swift+\nustar\ and \astrosat, despite the fact that the maximum seems to be at slightly lower energies in the \astrosat\ spectrum. The flux at the highest X-ray frequencies is, however, unchanged compared to previous observations. Similarly, the HE \g-ray spectra are comparable and connect well with the VHE spectrum. Therefore, only one spectrum is considered in the modelling.

For 1ES\,1741+196, we again collect three different spectra: Spectrum 1 (black in Fig.~\ref{fig:sseddt}) comprising of data from \swift, \fermi\ and MAGIC; Spectrum 2 (blue in Fig.~\ref{fig:sseddt}) containing data from \astrosat\ (A1+A2) and \fermi; and Spectrum 3 (red in Fig.~\ref{fig:sseddt}) with data from \astrosat\ (A3) and the same \fermi\ spectrum as for Spectrum 2.
There is noticeable flux and spectral variability in the X-ray domain indicating a flux rise from Spectrum 1 to Spectrum 2. In the few months passing to Spectrum 3, the X-ray peak frequency drops.
The optical/UV spectra seem to have a strong contribution from the host galaxy (hosted in a triplet of interacting galaxies, \cite{magic17}) and the emission is roughly stable from Spectrum 1 to 3. The spectral index from the optical/UV to the X-ray domain increases from Spectrum 1 to Spectra 2 and 3.
On the other hand, the \g-ray spectrum seems stable. Variations on monthly timescales, as in the X-ray domain, cannot be detected, unfortunately, due to the low flux, which is also why only one spectrum is shown for periods 2 and 3. The overall soft \g-ray spectrum leads to the classification as an \esyn\ source.

The HBL 1ES\,2322-409 does not show significant spectral variation in the X-ray band. Compared to \swift\ observations, a mildly higher flux is noticeable. The HE and VHE \g-ray spectra are well connected despite the significant time span between the observations, also suggesting stable fluxes. Because of this stability, we only consider one spectrum for this source.


\subsection{Constraints} \label{sec:modconstraints}

The shape of the spectral components in the SEDs provides important constraints on the particle distributions. We define the spectral index $\alpha$ in a given energy range as $\nu F_{\nu}\propto \nu^{\alpha}$. While the \fermi\ \g-ray spectrum directly gives us the \g-ray spectral index, $\alpha_{\g}$, we need to make an assumption on the shape of the synchrotron spectrum, as the X-ray spectra only resemble peaks and cut-offs, but no broad power law except for the HBL 1ES\,2322-409. On the other hand, it is plausible that below the X-ray domain the synchrotron spectrum resembles a power-law smoothly connecting to lower energies. Unfortunately, in several cases the UV data points are influenced by non-jetted emission (e.g., the host galaxy). While one can attempt a joint fit of the synchrotron power-law and the galactic components \citep[as in, e.g.,][]{Wierzcholska+2020}, it is a reasonable approximation to ignore this influence. In turn, the spectral index derived from the UV to X-ray spectrum can be considered as a lower limit, 
i.e., the synchrotron spectrum could be harder. The derived spectral indices of the optical/UV-to-X-ray range, $\alpha_{ox}$, and the HE \g-ray range, $\alpha_{\g}$, are given in Table~\ref{tab:modelsources}. 

Within errors, $\alpha_{ox}$ and $\alpha_{\g}$ agree for most sources and spectra.
Given that 1ES\,2322-409 is an HBL, the different values of $\alpha_{ox}$ and $\alpha_{\g}$ are expected. It is nonetheless assuring that for this source the index $\alpha_{ox}$ agrees well with the spectral index in the X-ray domain suggesting that the synchrotron peak is located in the optical/UV range.

The synchrotron spectral index, $\alpha_{\rm sy}$, is directly related to the 
spectral index of the injected/accelerated particle distribution,
$s$, through the relation $s = 3 - 2\alpha_{\rm sy}$ for uncooled particles, and $s = 2 - 2\alpha_{\rm sy}$ for cooled particles. While the cooling status must be verified \textit{a posteriori}, this can be used to constrain the electron spectral index from $\alpha_{ox}$. Similarly, if the \g\ rays stem from proton synchrotron emission, the proton distribution is directly given from $\alpha_{\g}$.


The lack of (observed) variability on time scales shorter than a few days prevents us from any meaningful constraint on the source size. We will thus follow \cite{cerruti+15}, and employ standard one-zone sizes on the order of $10^{15-17}\,$cm. These authors also showed an inverse relation between source size and magnetic field strengths keeping the Lorentz factor of the cooling break constant, and that a relatively large range of the parameter space can produce reasonable fits. Furthermore, small region sizes and high magnetic fields result in a lower overall source power. We thus concentrate on this parameter range.


The jet power is an important measure to quantify a model's plausibility beyond a fit to the data. As the jet is anchored in the black-hole-disk system, the jet power is tied to the power funneled through the accretion disk to the black hole. The accretion power is thus an important measure against which the jet power can be gauged. 
However, we have no direct observational evidence of the disk flux in any of our sources. In turn, we chose the accretion disk luminosities such that the summed flux of the disk and the jet does not overshoot the observed data. The obtained values are given in Tab.~\ref{tab:totalpower} and can be regarded as upper limits. We note that the employed radiation codes (see below) use standard Shakura-Sunyaev disks \citep{shakurasunyaev73}, while HBL and eHBL sources are typically regarded as hosting radiatively inefficient accretion flows  \citep[RIAFs; e.g.,][]{igumenshchev04}. This implies that the obtained luminosity limits on the disks in Tab.~\ref{tab:totalpower} may not be the true accretion power, as RIAFs can sustain much higher accretion rates than suggested by their emitted radiation \citep{katz77,czerny19,ghodlaeldrige23}. Nonetheless, the luminosity limit may still provide important constraints.

Similarly, the masses of the supermassive black holes are uncertain or even unknown in our sources. In order to provide references in the discussions below, we provide the Eddington luminosity for a black hole with mass $1\E{8(9)}\,M_{\odot}$ being $1.3\E{46(47)}\,$erg/s. 
In any case, the inferred limits on the radiation output of the accretion disks are orders of magnitude below the Eddington limit.

The powers in the observer's frame, $\hat{P}$, for the radiation, magnetic field, electron and proton population are calculated with

\begin{align}
    \hat{P}_i = \pi R^2 c \Gamma^2 u_i
    \label{eq:obspow},
\end{align}
with the bulk Lorentz factor $\Gamma$ and the energy density $u_i$ of the respective constituent. The energy density of the radiation, $u_{rad}$, is calculated from the model SED in the observer's frame, $\hat{\nu} \hat{F}_{\hat{\nu}}$, with the relation \citep{zw16}

\begin{align}
    u_{rad} = \frac{4d_L^2}{cR^2\delta^4}\int \hat{\nu} \hat{F}_{\hat{\nu}} \td{\ln{(\hat{\nu})}}
    \label{eq:urad}.
\end{align}
The magnetic energy density is $u_B=B^2/8\pi$, while the particle energy densities are given as

\begin{align}
    u_{e/p} = m_{e/p}c^2 \int \gamma n_{e/p}(\gamma) \td{\gamma}
    \label{eq:upart}.
\end{align}
There are two caveats here. First, in the \SSC\ model we assume one cold proton per electron giving the proton energy density as $u_p = m_pc^2\int n_e(\gamma)\td{\gamma}$. Second, the proton power in the other models depends strongly on the minimum proton Lorentz factor, $\gamma_{{\rm min},p}$, which we assume close to unity owing to the lack of constraints. Larger values of $\gamma_{{\rm min},p}$ could reduce the proton power substantially.
The total jet power, $\hat{P}_{\rm jet}$, is the sum of the four constituents and is listed for each source and model in Tab.~\ref{tab:totalpower}. The individual powers are given in App.~\ref{app:tables}.

%
\subsection{Modeling} \label{sec:modeling}
We use various codes to model the SEDs of our sources. Here, we only describe the purpose of the codes and the results, while brief code descriptions including definitions of the free parameters are provided in appendix \ref{sec:codes}. In all cases, the model curves have been derived as fits-by-eye, as a broad range of solutions is possible in all cases \citep[e.g.,][]{cerruti+15}. Steady-state solutions have been obtained for all SEDs given the lack of variability information on short timescales as well as the non-simultaneity of the data. 

Firstly, we derive a simple leptonic one-zone SSC model (hereafter referred to as \SSC) using the steady-state code of \cite{Boettcher13}. In the plots of the SED fits, this model is shown as the red solid line. 

Secondly, we use the electron-proton co-acceleration model (hereafter referred to as \epshock) of \cite{Zech2021}. The advantage of this model is a physical motivation of the hard electron distribution. However, as this model is specifically designed to explain hard intrinsic VHE spectra, it is only applied to the \eTeV\ sources 1ES\,0120+340, RGB\,J0710+591, and 1ES\,1101-232. 
A magenta dash-double-dotted line marks this model in the SED-fit plots. 

Thirdly, we employ the lepto-hadronic code \onehale\ \citep[version 1.1,][]{Zacharias21,zacharias+22}. We produce two solutions with this code. The first one is a lepto-hadronic model (hereafter referred to as \lhpi), where the \g\ rays are produced from electron-synchrotron emission by secondary pairs (from Bethe-Heitler pair production, photo-pion production, and $\gamma\gamma$ pair production). This is actually a specific choice, where we suppress the SSC contribution, which can be prominent in an \lhpi\ model \citep[cf.,][]{cerruti+15}. Blue dashed lines show this model. The second solution is a proton-synchrotron model (hereafter referred to as \lhp) in order to describe the \g\ rays, which is displayed as orange dash-dotted lines.

Below, we discuss each source in turn providing individual SED fits. The complete sets of model parameters are given in App.~\ref{app:tables}. 
Given the large number of free parameters, especially in the lepto-hadronic models, we try to keep as many parameters as possible fixed from model to model as well as from source to source. This includes the Doppler factor, $\delta$, the escape and acceleration time parameters, $\eta_{\rm esc}$ and $\eta_{\rm acc}$, and, in the lepto-hadronic models, the magnetic field $B$. 
For instance, we fix the Doppler factor to $\delta=50$ in all cases and also employ $\delta=\Gamma$. 
While this is a large value, and sometimes better fits are possible with lower values, it removes an ambiguity between the models and eases the interpretation.
In turn, the main differences in the modeling arise from the particle distributions and the size of the emission region.

\subsubsection{1ES\,0120+340} \label{sec:mod0120}
\begin{figure}
\centering
\includegraphics[width=0.48\textwidth]{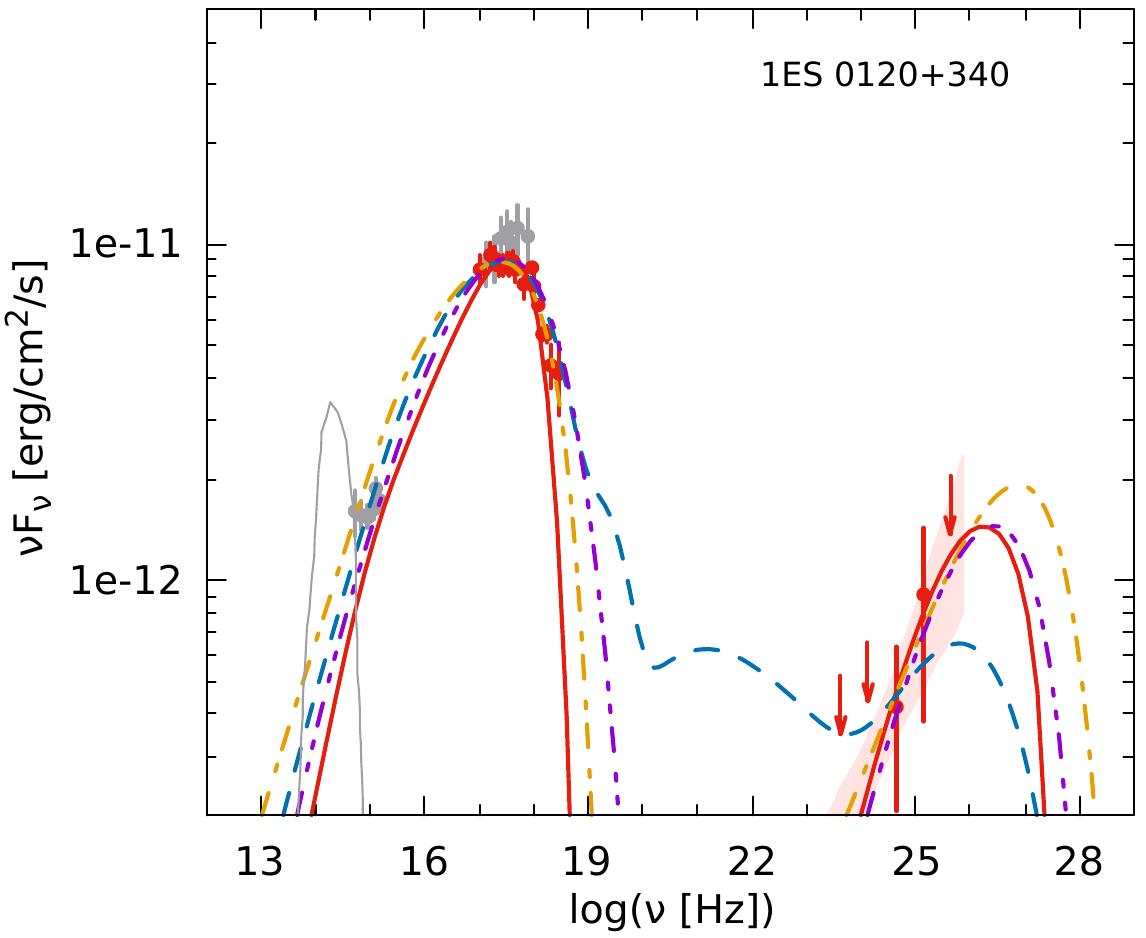}
\caption{Main (red) and archival (gray) data sets of 1ES\,0120$+$340, as well as the various intrinsic models: \SSC\ (red solid line), \epshock\ (magenta dash-double-dotted line), \lhpi\ (blue dashed line), and \lhp\ (orange dot-dashed line). 
The thin gray line marks the host galaxy template of \protect\cite{silva+98}.
}
\label{fig:models0120} 
\end{figure}
%

The fits to 1ES\,0120+340 are shown in Fig.~\ref{fig:models0120}, while the model parameters are given in Tab.~\ref{tab:models0120}.
%
The fits to the data are generally good. Differences occur in the VHE \g-ray domain, which could become an important discriminator should this source be established as a VHE source in future observations. 
Modelling 1ES\,0120+340 with the \epshock\ model assuming acceleration on a single shock did not yield a satisfactory result. Allowing for re-acceleration on a second shock provides us with a good fit.
The \lhpi\ model results in a flat spectrum above $10^{20}\,$Hz. The model MeV bump is synchrotron emission of Bethe-Heitler-pair-produced electrons, while the GeV bump is synchrotron emission of electrons from \g-\g\ pair production and pion production. No VHE \g-ray emission is expected from this model.

The particle spectral indices in the \SSC\ and \lhpi\ models suggest that the particles are cooled. In the \lhp\ case, this is only true for the electrons, while the protons are not cooled. This requires a softer proton injection distribution compared to the electrons. We point out here that in the \epshock\ model the hardening of the electron distribution after injection due to (additional) acceleration is taken into account. Thus, the consideration concerning the injection spectral index of the particle distributions derived from the observed spectra does not apply.

The other parameters are comparable with the other sources below with no significant outlier. It is interesting, though, that the optical -- X-ray and HE \g-ray spectra are among the hardest ones in our list of sources requiring very hard particle injection distributions.

In all models, the jet power is particle-dominated. All total jet powers exceed the accretion disk luminosity limit derived from the modeling. While the \SSC\ model and the \epshock\ model are within an order of magnitude of the disk limit, the \lhpi\ and \lhp\ models exceed the limit by several orders of magnitude. The \lhpi\ model even exceeds the Eddington luminosity of a $10^9\,M_{\odot}$ black hole.


%
\subsubsection{RGB\,J0710+591} \label{sec:mod0710}
\begin{figure}
\includegraphics[width=0.48\textwidth]{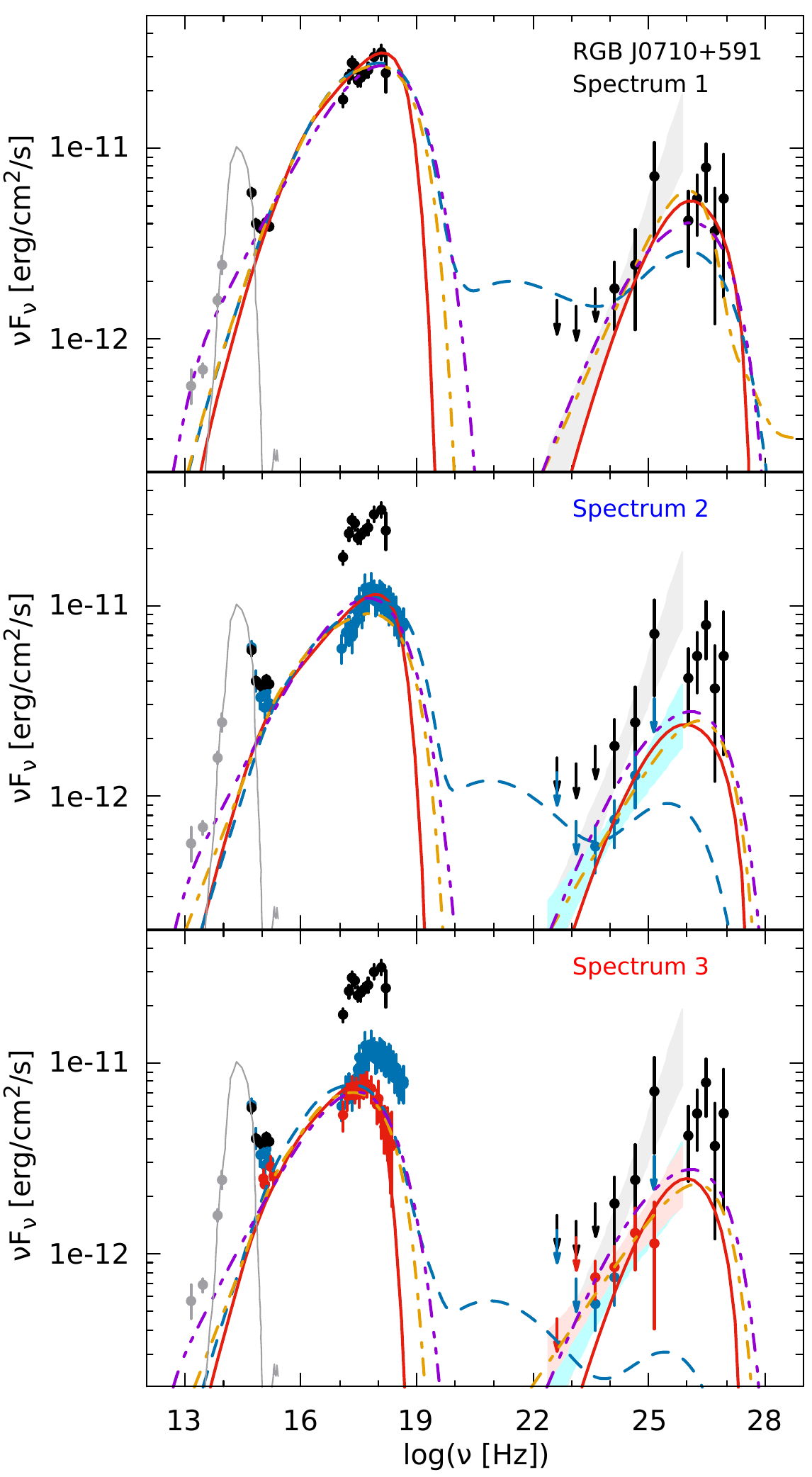}
\caption{Main (color) and archival (gray) data sets of RGB\,J0710$+$591, as well as the various intrinsic models [\textit{Top:} for Spectrum 1 (black); \textit{Middle:} for Spectrum 2 (blue); \textit{Bottom:} for Spectrum 3 (red)]: \SSC\ (red solid line), \epshock\ (magenta dash-double-dotted line), \lhpi\ (blue dashed line), and \lhp\ (orange dot-dashed line). 
VHE \g-ray data have been corrected for EBL absorption.
The thin gray line marks the host galaxy template of \protect\cite{silva+98}.
}  
\label{fig:models0710} 
\end{figure}
%

The three SEDs shown in Fig.~\ref{fig:models0710} indicate significant changes between them. 
Spectrum 1 exhibits the highest X-ray flux, as well as the hardest HE \g-ray spectrum. In fact, judging from Fig.~\ref{fig:j0710_lc}, RGB\,J0710+591 was in a prolonged HE high-state during this time with a subsequent flux decrease. This decrease is accompanied by a softening of the HE spectrum. Unfortunately, no data of the VHE \g-ray spectrum exists for the later epochs, but a flux drop is likely along with the softening of the HE spectrum --- even though, to be clear, the VHE spectrum might still be consistent with an extension of the later HE spectra within statistical errors. The X-ray spectrum drops in flux and seemingly exhibits spectral changes in the later data sets. While the first X-ray spectrum is compatible with a pure power-law with $\alpha_X=0.25$, the second spectrum clearly indicates a curved spectrum with a peak below $10\,$keV, which further drops in Spectrum 3. However, we cannot rule out the presence of such a peak in Spectrum 1 given the limited spectral coverage of \swift-XRT. Interestingly, the optical-X-ray spectra in the second and third data sets are much softer than the first one suggesting that the underlying electron distribution has softened. 
The parameter sets for the three spectra are given in Tabs.~\ref{tab:models0710old1}, \ref{tab:models0710old2} and~\ref{tab:models0710}, respectively.
%

With the exception of the \lhpi\ model, the fits are good for the various models in all three states. The poor \lhpi\ model fit is due to the constraint to be consistent with the upper limits at the lowest \g-ray energies, which makes it impossible to reproduce the subsequent hard \g-ray spectrum up to the VHE domain.

In the \SSC\ model and the \lhpi\ model, the particle distributions are cooled, while this is only true for the electrons in the \lhp\ model. In the latter, the protons are uncooled leading to a softer injection distribution compared to the electrons. This is true for all three source states. 

In order to accommodate the changes between the data sets, relatively minor changes must be done from data set to data set.
In the \SSC\ model, the main change is in the magnetic field which drops from 0.03\,G to 0.02\,G and 0.015\,G. An increase in the electron power is required from Spectrum 1 to Spectrum 2 in order to account for the slightly rising peak-flux ratio between the low- and the high-energy components of the SED. The power drops then to Spectrum 3, and the maximum electron Lorentz factor is reduced.
Generally, the parameters are consistent with the modeling of \cite{veritas10} and \cite{Costamante+2018}. The parameters of \cite{katarzynski12} differ from our estimates as he used a much lower Doppler factor of $\delta=8$.
In the \epshock\ model, we find a continuous increase of the radius and a continuous decrease of the magnetic field from spectrum to spectrum. However, the electron distribution does not change from Spectrum 1 to Spectrum 2, and only reduces mildly in energy density to Spectrum 3. The energy density of the proton distribution (which is important for the electron acceleration) drops continuously from Spectrum 1 to 3. The behaviour is reminiscent of adiabatic expansion of the emission region \citep{boulamastichiadis22,zacharias23}, but the magnetic field strength varies too rapidly with respect to the radius and the overall time-scale is much too long to be explained with the relativistic movement of a blob along the jet.
%
As we keep the magnetic field constant in the \lhpi\ model, the spectral changes are mostly accounted for through a reduction of the electron injection power, as well as shifts in the minimum and maximum electron Lorentz factors. In order to produce the secondary electron population, the proton distribution has to be changed in non-trivial manners. In order to accommodate the reduced upper limits in Spectrum 2 compared to Spectrum 1, the maximum proton Lorentz factor must be reduced to shift the cut-off in the Bethe-Heitler component (the peak at $\sim 10^{21}\,$Hz) to lower energies. However, this requires an increase in the proton power to achieve the flux of the upper limits. The proton power is reduced again in order to account for Spectrum 3.
In comparison to the modeling in \cite{cerruti+15}, we employ a smaller emission region and a higher magnetic field in order to suppress the SSC contribution, which is used in \cite{cerruti+15}. This choice also has consequences for the proton distribution, which in their case has a higher maximum proton Lorentz factor and a lower proton power than in our modeling.
The \lhp\ model requires more important adjustments from case to case due to the change in the HE \g-ray domain. The softening of the HE spectrum is best reproduced by a softening of the proton distribution plus an increase in the magnetic field. The latter shifts the synchrotron spectrum to higher energies, allowing for an improved representation of the \g-ray data. In addition, the proton power must be increased considerably to counter the flux reduction at the highest energies due to the softening of the proton distribution. The increase in the magnetic field requires a significant reduction in the minimum and maximum electron Lorentz factors from Spectrum 1 to Spectrum 2 along with the drop in particle power.
The parameter sets are generally in the range obtained by \cite{cerruti+15}.

In all our cases, the jet power is particle dominated and exceeds the inferred upper limit of the accretion disk. 
Interestingly, for all models the jet power does not decrease from state to state as one would expect from the observed flux drops (see Tab.~\ref{tab:totalpower}). The \SSC\ model barely requires any change in jet power, while in the the \epshock\ model the power even increases as the decrease in magnetic field strength is countered by an increase in the particle power. In the \lhpi\ model, Spectrum 2 exhibits the highest jet power due to the required increase in proton power. In all three states, the jet power in the \lhpi\ model surpasses the Eddington power of a $10^9 \,M_{\odot}$ black hole. The aforementioned increase of the proton power in the \lhp\ model induces an increase in jet power similar to the \epshock\ model. For Spectrum 1, the \lhp\ jet power is below the Eddington luminosity of a $10^8 \,M_{\odot}$ black hole, while it surpasses that of a $10^9 \,M_{\odot}$ black hole in Spectrum 3.\footnote{We made the generic comparisons for consistency with the other sources. However, RGB\,J0710+591 is the only source in our sample with a mass estimate of the black hole: $log{M_{\rm BH}} = 8.25 \pm 0.22$ \citep{2005ApJ...631..762W}. The corresponding Eddington luminosity is $L_{\rm Edd}=2.34\E{46}\,$erg/s.}


%
\subsubsection{1ES\,1101-232} \label{sec:mod1101}
\begin{figure}
\includegraphics[width=0.48\textwidth]{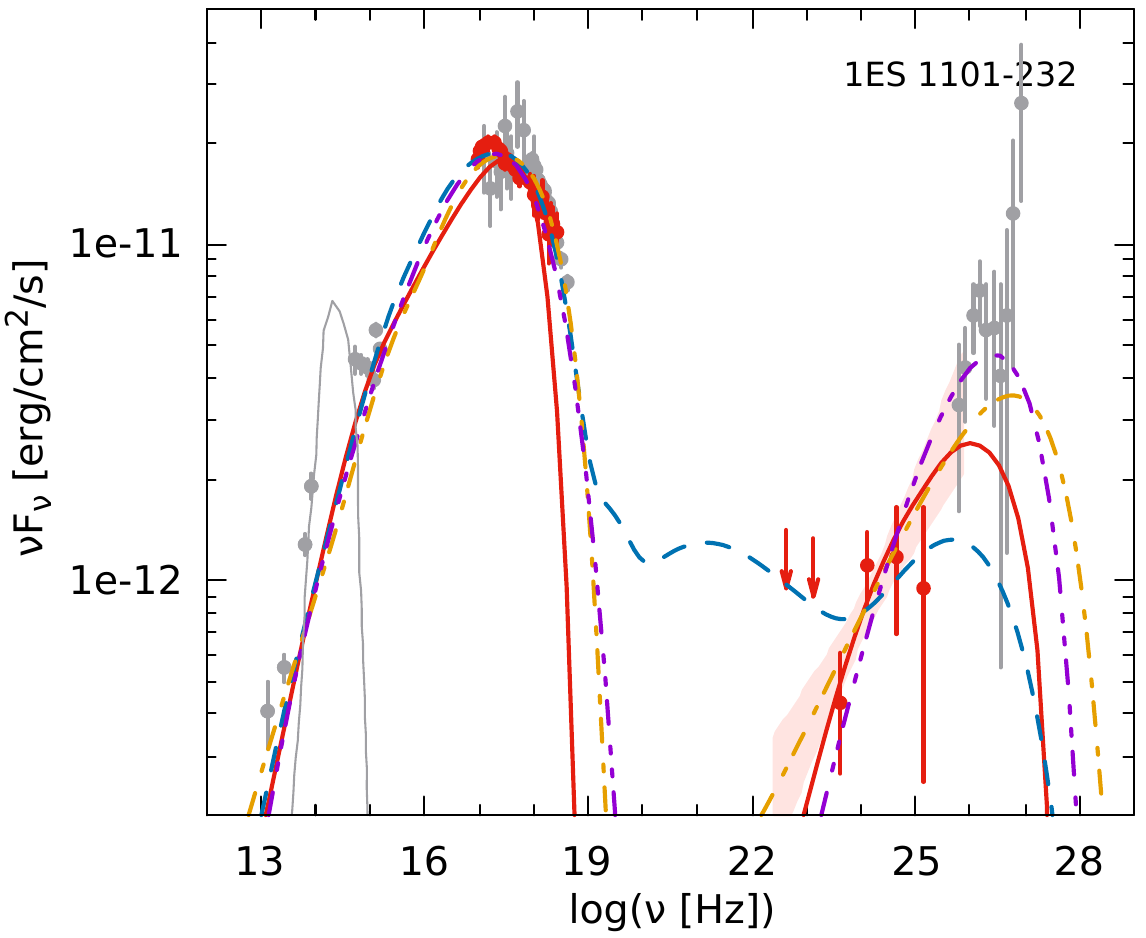}
\caption{Main (red) and archival (gray) data sets of 1ES\,1101$-$232, as well as the various intrinsic models: \SSC\ (red solid line), \epshock\ (magenta dash-double-dotted line), \lhpi\ (blue dashed line), and \lhp\ (orange dot-dashed line). 
VHE \g-ray data has been corrected for EBL absorption.
The thin gray line marks the host galaxy template of \protect\cite{silva+98}.
}  
\label{fig:models1101} 
\end{figure}
%

The fits for this source are displayed in Fig.~\ref{fig:models1101}, while the parameters can be found in Tab.~\ref{tab:models1101}.
%
The \epshock\ model and the \lhp\ model reproduce the data well, while the \SSC\ model and the \lhpi\ model cannot reproduce the (archival) VHE data. Additionally, the \lhpi\ model does not work well for the low-energy part of the HE \g-ray spectrum.

The particle distributions in the \SSC\ model and the \lhpi\ model are cooled, which also holds for the electrons in the \lhp\ model. The protons in the \lhp\ model are, however, uncooled, resulting in a softer injection distribution than the electrons.

The jet power is particle dominated in all cases, and all models surpass the inferred upper limit of the accretion disk luminosity. The \lhpi\ model exceeds the Eddington power of a $10^9 \, M_{\odot}$ black hole, while the \lhp\ model requires at least a black hole of mass $7\E{8}\,M_{\odot}$ in order to remain sub-Eddington.

This source has also been modeled by other authors.
\cite{hess07} and \cite{Costamante+2018} employed SSC models. Given the difference in the Doppler factors used between each other and with respect to our modeling, the difference in the other parameters is obvious. 
The \epshock\ model was already considered for this source in \cite{Zech2021}. Compared to their work, we require a larger emission region and smaller particle density and magnetic field strength, given a slightly different spectral shape in the X-ray domain. Particle acceleration must occur at two shocks in order to achieve a reasonable fit. 
%
\cite{cerruti+15} produced lepto-hadronic models for this source.\footnote{Their HE \g-ray spectrum is rather soft, as the spectrum from the 2FGL catalog was used, which had very limited statistics.} While their \lhpi\ model fits the VHE data, it also suggests a significant Bethe-Heitler component, which would overwhelm our HE \g-ray spectrum even more than our \lhpi\ model already does. This suggests that this model is indeed not a good solution for this source.
Their \lhp\ model parameters are very similar to ours.


%
\subsubsection{1ES\,1741+196} \label{sec:mod1741}
\begin{figure}
\includegraphics[width=0.48\textwidth]{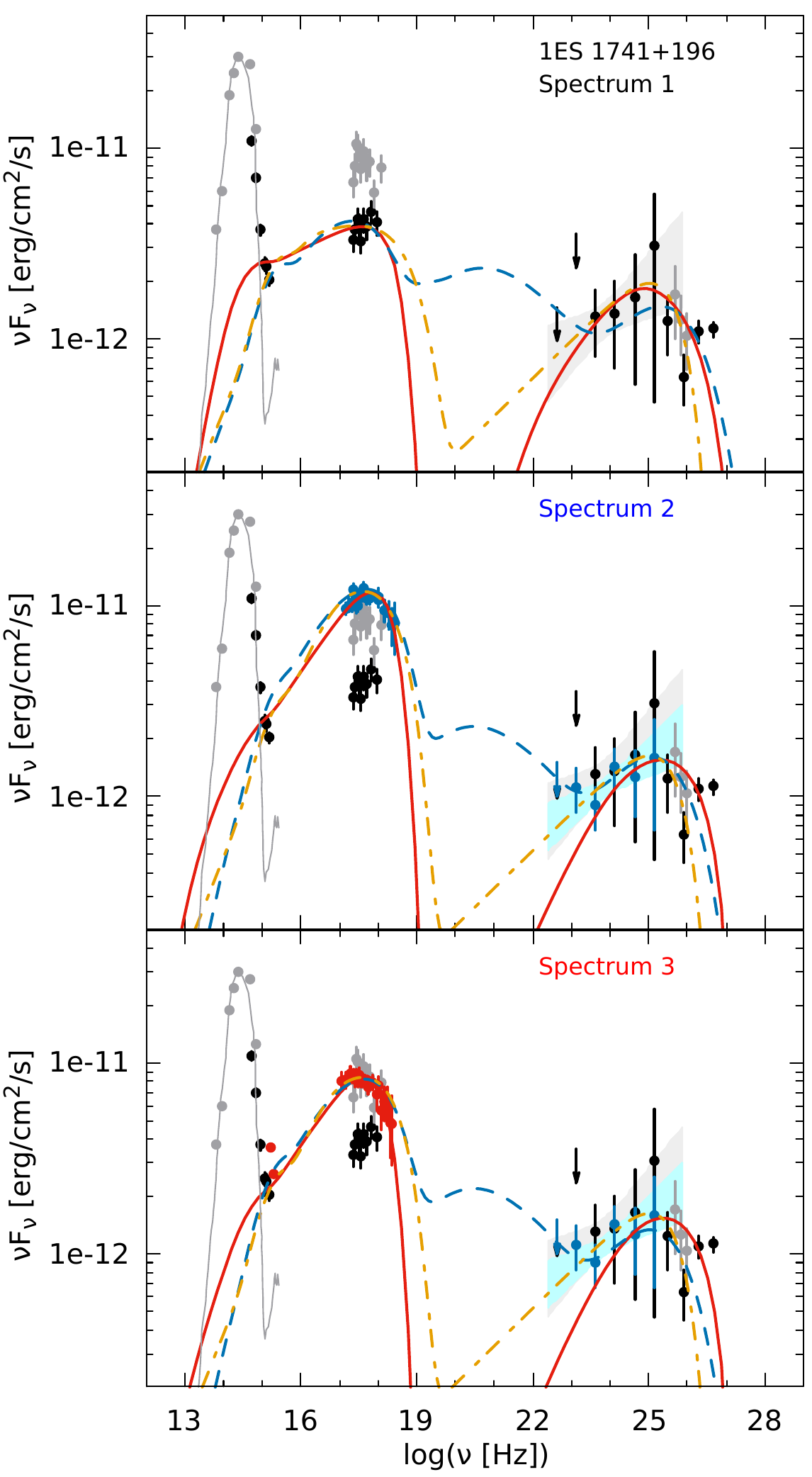}
\caption{Main (color) and archival (gray) data sets of 1ES\,1741$+$196, as well as the various intrinsic models [\textit{Top:} for Spectrum 1 (black); \textit{Middle:} for Spectrum 2 (blue); \textit{Bottom:} for Spectrum 3 (red)]: \SSC\ (red solid line), \lhpi\ (blue dashed line), and \lhp\ (orange dot-dashed line). 
VHE \g-ray data has been corrected for EBL absorption.
The thin gray line marks the host galaxy template of \protect\cite{silva+98}.
}  
\label{fig:models1741} 
\end{figure}
%

Due to its flat HE \g-ray spectrum, we categorize 1ES\,1741+196 as an \esyn\ source. Since the \epshock\ model has been set up specifically to describe the very narrow spectral bumps of extreme TeV blazars, it is not applicable to this source in its current form. Nonetheless, the source shows an interesting MWL evolution from state to state.
Spectrum 1 indicates a soft optical/UV -- X-ray spectrum with $\alpha_{ox}\sim 0.1$, implying a soft underlying electron distribution with $s_e\sim 1.8$. Surprisingly, the hardening of the synchrotron spectra between the two \astrosat\ observations is not reflected in the HE \g-ray spectrum. This complicates the interpretation. 
The models are shown in Fig.~\ref{fig:models1741}, while the parameters are listed in Tabs.~\ref{tab:models1741old}, \ref{tab:models1741a}, and~\ref{tab:models1741b}.

The three applied models reproduce all three data sets fairly well.
%
In the \SSC\ model and the \lhpi\ model, the particle distributions are cooled. In the \lhp\ model, the proton distribution is uncooled, while the electron distribution is cooled. In turn, the proton injection distribution is softer than the electron distribution.

As mentioned above, the X-ray spectrum shows significant variability while the \g-ray spectrum remains steady. From Spectrum 1 to Spectrum 2, the X-ray flux rises by at least a factor 3 with a mild subsequent drop to Spectrum 3. The X-ray peak frequency does not seen to shift significantly from state to state, even though a clear determination of the peak in Spectrum 1 is not possible.
In order to reproduce these changes with the \SSC\ model, the most important change is a higher particle power in addition to a change in the particle spectral index. The radius and the magnetic field change by at most $50\%$. The escape time factor, $\eta_{\rm esc}=400$, is very high. An SSC model was also employed by \cite{veritas16} and \cite{magic17}. However, their X-ray spectra do not show the cut-off that we see, especially with the \astrosat\ data. Therefore, our electron energy distribution is much more restricted. As we use a higher Doppler factor than either of those earlier works, the differences in radius (ours being smaller) and magnetic field (ours being higher) are reasonable.
In the \lhpi\ model and the \lhp\ model, the variations in our spectra can be reconciled easily with minor changes of at most a factor $2$ in the parameters. 
%

In all our cases, the total jet power is dominated by particles and surpasses the inferred upper limit of the accretion disk luminosity. Both the \lhpi\ model and the \lhp\ model exceed the Eddington luminosity of a $10^9 \, M_{\odot}$ black hole in all 3 data sets owing to the soft proton distribution requiring a large power.


%
\subsubsection{1ES\,2322-409} \label{sec:mod2322}
\begin{figure}
\includegraphics[width=0.48\textwidth]{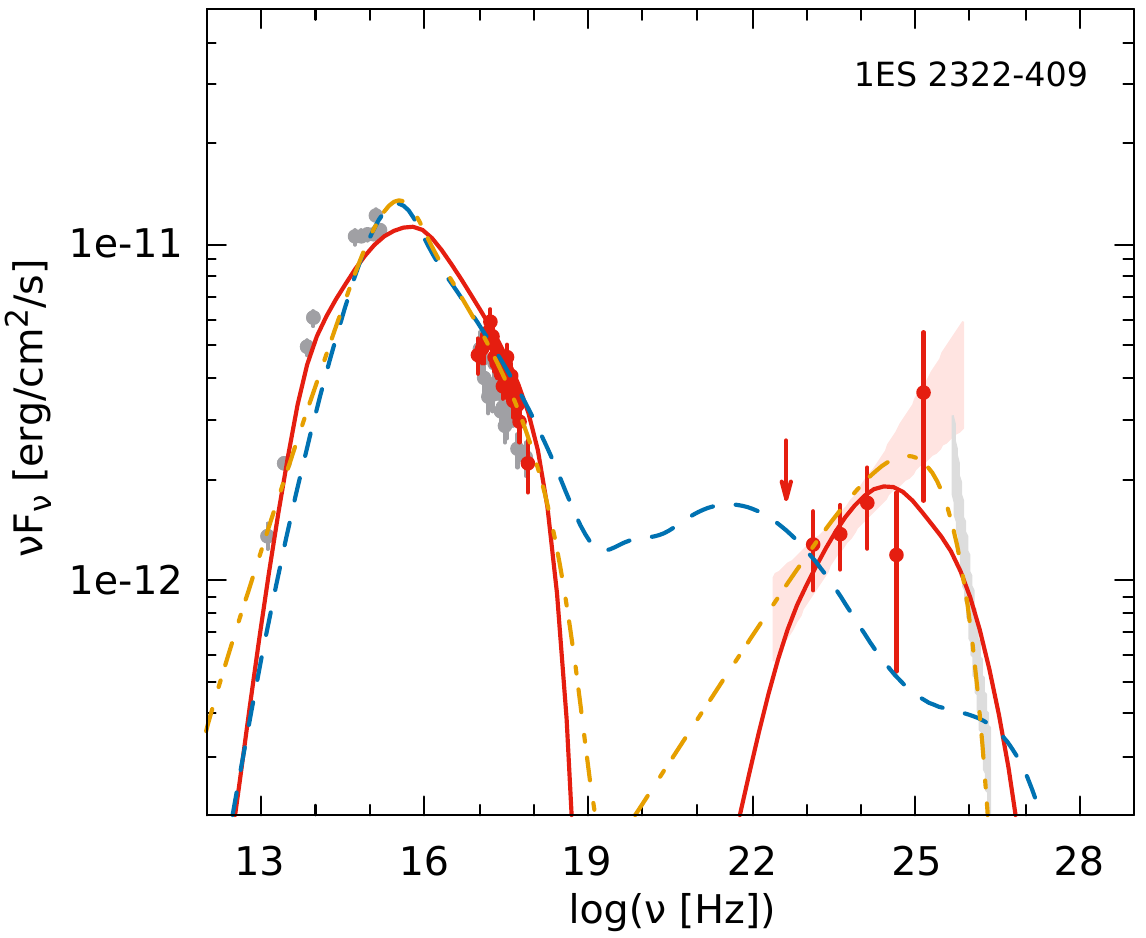}
\caption{Main (red) and archival (gray) data sets of 1ES\,2322$-$409, as well as the various intrinsic models: \SSC\ (red solid line), \lhpi\ (blue dashed line), and \lhp\ (orange dot-dashed line). 
VHE \g-ray data has been corrected for EBL absorption.
}  
\label{fig:models2322} 
\end{figure}
%

This is a classical HBL source with a soft X-ray spectrum. The \epshock\ model is, thus, not applied here either. 
The fits are displayed in Fig.~\ref{fig:models2322}, while the parameters are given in Tab.~\ref{tab:models2322}.

The \SSC\ model and the \lhp\ model fit the data very well, while we were not able to find an acceptable fit for the \lhpi\ model. The reason is the low synchrotron peak energy and, in turn, the soft X-ray spectrum. As these are the target photons for the p\g\ interactions, even a very hard proton distribution would not produce a significant secondary flux from pion decay needed in the HE domain. Additionally, due to the soft X-ray target photon field, \g\g\ pair production also contributes very little to the HE domain. These two effects diminish the synchrotron peak at HE \g\ rays, which is produced by secondary electrons from these two interaction channels.

The electron distributions in all models are cooled, while the proton distributions in the \lhpi\ and \lhp\ models are uncooled. Nonetheless, for this source this implies that in the lepto-hadronic models the spectral indices of the electron and proton injection distributions are equal.

The remaining parameters are well in line with parameters for other HBL sources. This is best exemplified by the fact that the parameters of our \SSC\ model agree very well with the parameter set of \cite{hess19}.

The total jet power is particle dominated in all cases. While the \SSC\ model is only barely exceeding the inferred upper limit of the accretion disk luminosity, the \lhpi\ model and the \lhp\ model exceed this limit by orders of magnitude and even exceed the Eddington luminosity of a $10^9 \, M_{\odot}$ black hole.

Even though this source shows peak frequencies that are more usual for regular HBLs, the low value of the magnetic field strength and high value of the minimum electron Lorentz factor required for the \SSC\ model, indicate some common features with extreme TeV blazars.  


%
%
\begin{table*}
\centering
\footnotesize
	\caption {Total jet powers $\hat{P}_{\rm jet}$ in erg/s for the various model fits, as well as the upper limit on the accretion disk power from the modeling. In case of multiple source spectra, the accretion disk power is not changed from case to case.} 
\begin{tabular}{lc|cccc|c}
\hline \hline 
	Source & Spectrum & \SSC\ & \epshock\ & \lhpi & \lhp & Acc. Disk \\
\hline 
1ES\,0120$+$340 &  1 & $7.7\E{43}$ & $1.6\E{45}$ & $1.6\E{48}$ & $3.5\E{45}$ & $3\E{43}$ \\
\hline
RGB\,J0710$+$591 & 1 & $3.1\E{43}$ & $6.4\E{44}$ & $5.5\E{47}$ & $3.2\E{45}$ & $1.1\E{43}$ \\
 & 2   & $3.2\E{43}$ & $1.0\E{45}$ & $1.1\E{48}$ & $3.3\E{46}$ &  \\
 & 3   & $3.4\E{43}$ & $1.8\E{45}$ & $5.6\E{47}$ & $2.4\E{47}$ &  \\
\hline
1ES\,1101$-$232 &  1 & $9.3\E{43}$ & $2.6\E{45}$ & $9.8\E{47}$ & $8.7\E{46}$ & $4.7\E{43}$ \\
\hline
1ES\,1741$+$196 & 1  & $1.4\E{43}$ & -- & $9.0\E{48}$ & $8.3\E{48}$ & $1\E{43}$ \\
 & 2   & $1.2\E{43}$ & -- & $3.3\E{48}$ & $6.9\E{48}$ &  \\
 & 3   & $1.3\E{43}$ & -- & $4.4\E{48}$ & $6.9\E{48}$ &  \\
\hline
1ES\,2322$-$409 & 1  & $1.2\E{44}$ & -- & $6.1\E{49}$ & $2.1\E{48}$ & $1\E{44}$ \\
\hline \hline 
\end{tabular} \\
\label{tab:totalpower}
\end{table*}
\section{Discussion \& Conclusions} \label{sec:summary}


In this paper, we have presented  the data analysis of three \eTeV sources, one \esyn\ and one HBL source observed with \astrosat\ and other instruments. For the first time, we established the X-ray peak energy in two sources, namely 1ES\,0120+340 and 1ES\,1741+196. The former source exhibits \eTeV\ characteristics and is thus a VHE-\g-ray-detection candidate. A VHE \g-ray detection would strongly constrain the model-parameter space. Furthermore, while 1ES\,0120+340, and 1ES\,1102-232 did not show any variability compared to archival data sets, clear variability is established in RGB\,J0710+591 and 1ES\,1741+196. 
The HBL 1ES\,2322-409 does not show variability in our data sets, however it is known to be variable at least in the synchrotron component \citep{hess19}.

RGB\,J0710+591 exhibits both flux and spectral variability in the X-ray and \g-ray bands. The X-ray long-term light curve (Fig.~\ref{fig:j0710_lc}) shows variations on days to weeks time scales, while a marginally significant high state in the \g-ray band is visible in the first years of observations. The long-lasting downward trend in the X-ray light curve over the following few years of observations is accompanied by a drop in the X-ray peak frequency (Fig.~\ref{fig:sseddt}). On similar time scales, the \g-ray spectrum softens. Given that neither of the \fermi\ spectra connects well to the archival TeV spectrum, the latter energy range might also be variable. Additional observations in that domain should clarify this point. The reproduction of the changes requires non-trivial parameter changes depending on the chosen model as described in Sec.~\ref{sec:mod0710}. While none of the solutions is unique, and different parameter sets might provide equivalent fits, these results suggest that there is no simple physical explanation for these changes.

On the contrary, 1ES\,1741+196 shows variability mainly in the X-ray domain, while there is no obvious variability in the optical and \g-ray regimes. The former implies that the electron distribution has to change from one spectrum to the next in order to accommodate the relative change between the optical and X-ray fluxes. This actually makes it complicated to account for the non-changing \g-ray spectrum in a leptonic model, which is reflected in the way the parameters have to be changed. This is the advantage of the \lhpi\ and \lhp\ models, as the proton distribution does not need to be changed. However, the power demand of the lepto-hadronic models is a problem.


Indeed, the modeling results highlight the pros and cons of the various models. The \SSC\ model is clearly the most conservative in terms of power requirements, but it has some issues with reproducing the full \g-ray spectra in the eHBL sources. The \lhp\ model has no issue with reproducing spectra owing to its large number of free parameters, but it has a huge power demand. Similarly, the \lhpi\ model requires a large amount of power and additionally has problems properly reproducing the \g-ray spectra. This is related to the fact that we specifically suppressed the SSC contribution in the \lhpi\ model resulting in an almost flat synchrotron SED of the secondary electrons. Fits with the \lhpi\ model improve when allowing for an SSC contribution \citep[e.g.,][]{cerruti+15}. However, this does not reduce the power demand.
All three models suffer from the fact that they are not designed to self-consistently explain the very hard particle injection distributions. The \lhp\ model has the additional complication that the proton and electron injection distributions do not exhibit the same power-law index. However, one would expect that the injection distributions from electrons and protons achieve more or less the same spectral index, if they were accelerated at the same shock or through the same process. 

This is the benefit of the \epshock\ model, which explains naturally the hard electron distributions as due to electron-proton-coacceleration at (multiple) shocks. However, by design it only works in \eTeV\ sources with a hard intrinsic VHE spectrum. In sources with a softer VHE spectrum (like \esyn\ sources and classical HBLs), the model is not directly applicable. Additionally in this scenario, one has to assume a large increase of about $10^2$ to $10^3$ in the magnetization between the upstream and downstream regions in \eTeV\ sources, while the upstream magnetization is rather low (order of $10^{-6}$). As pointed out by \cite{Tavecchio2022}, such a low magnetization may be a problem when ascribing the shocks to recollimation, since 3D MHD simulations indicate that only a single recollimation shock appears for sufficiently low magnetization due to instabilities that induce turbulence in the jet downstream of the recollimation shock. However, the amplification of the magnetization in the downstream region due to the particle stream might have an impact on the growth of turbulence. Other factors like jet stratification or the structure of large-scale magnetic fields might also play a role. 

As is typically observed for extreme blazars, all the objects in our sample require low magnetic fields (order of $10\,$ mG, except for 1ES\,1741+196) and high minimum electron Lorentz factors ($> 10^3$) in the \SSC\ and \epshock\ models. The HBL 1ES\,2322-409 shares these characteristics with the {\it bona fide} extreme blazars.

The jet power in all our models (including \SSC\ and \epshock) is above the inferred upper limits of the accretion disk luminosities. While in the \SSC\ model this may be due to the parameter choice, such a result is in line with the conclusion of \cite{ghisellini+14}, who found this to be a general feature in (high-power) blazars by comparing the observed \g-ray luminosity with inferred accretion disk luminosities. 
As we mentioned above, (e)HBLs are probably powered by RIAFs suggesting that the inferred radiation limits underpredict the true accretion power. Nevertheless, it is remarkable that the inferred limits on the disks are on the order of one per mill of the Eddington luminosity of even a $10^8\,M_{\odot}$ black hole.

In summary, the \astrosat\ and MWL observations have shown that extreme blazars exhibit various characteristics. While some of them are stable, others are variable on yearly, but also shorter, time scales. And while RGB\,J0710+591 varied in both X-rays and \g-rays, 1ES\,1741+196 only varied in the X-ray domain. The modeling suggests a preference of leptonic models due to the power demand -- even though all of our models exceed the obtained upper limits of the accretion disk luminosity, which is a curious fact. Given that lepto-hadronic models seem unlikely, we do not expect neutrinos to be emitted by these sources. In addition, neutrino production requires much more intense photon fields than are present in these sources \citep{Reimer2019}.

A study like ours would significantly benefit from long-term VHE \g-ray observations 
by Cherenkov telescopes. Unfortunately, no such data exists at the moment. As VHE \g-rays probe the high-energy peak of extreme blazars, they provide vital clues for the modeling, as well as valuable insights into the source characteristics and~--- especially~--- their variability. Proper MWL campaigns lasting several years will be crucial to gain more rigorous insights~--- not just about the sources themselves, but also about related studies such as probing intergalactic magnetic field \citep{hess2023}.


%
%
\section*{Acknowledgement}
The \textit{OneHaLe} code is available upon reasonable request to M.~Zacharias. 
This research has used the data of AstroSat mission of the Indian Space Research Organization (ISRO), archived at the
Indian Space Science Data Centre (ISSDC). This work has used the data from the Soft X-ray Telescope (SXT) developed at TIFR, Mumbai, and the SXT POC at TIFR is thanked for verifying and releasing the data via the ISSDC data archive and providing the necessary software tools. We acknowledge the High Energy Astrophysics Science Archive Research Center (HEASARC), which is a service of the Astrophysics Science Division at NASA/GSFC for providing data and/or software. This research has made use of the XRT Data Analysis Software (XRTDAS) developed by the ASI Space Science Data Center (SSDC, Italy) and NuSTAR Data Analysis Software (NuSTARDAS) jointly SSDC, Italy and the California Institute of Technology (Caltech, USA). The Fermi Science Support Center (FSSC) team is acknowledged for providing the data and analysis tools. M.~Zacharias acknowledges funding by the Deutsche Forschungsgemeinschaft (DFG, German Research Foundation) -- project number 460248186 (PUNCH4NFDI). I.~Sushch acknowledges support by the National Research Foundation of South Africa (Grant Number 132276).

%
%

\bibliographystyle{aa}
\bibliography{ehbl}

\begin{appendix}
\section{Code description} \label{sec:codes}

In this section we provide brief overviews of the codes used for the modeling of our sources.

\subsection{Steady-state leptonic one-zone model} \label{sec:boettcher13}

The code was developed by \cite{Boettcher13}, and calculates the radiative output of an electron distribution $n_e(\gamma)$ in a homogeneous, spherical region with radius $R$ permeated by a tangled magnetic field $B$. The electrons with Lorentz factor $\gamma$ obey the steady-state kinetic equation

\begin{align}
    -\frac{\pd}{\pd{\gamma}}\left[ |\dot{\gamma}(\gamma)|n_e(\gamma) \right] + \frac{n_e(\gamma)}{t_{\rm esc}} = Q_0 \gamma^{-s} \SF{\gamma}{\gamma_{\rm min}}{\gamma_{\rm max}}
    \label{eq:b13_kineq}.
\end{align}
The cooling term $\dot{\gamma}(\gamma)$ contains various radiative cooling terms, however in our case the most important ones are synchrotron and SSC cooling. The escape time scale is given as a multiple of the light-crossing time scale: $t_{\rm esc}=\eta_{\rm esc}R/c$ with $\eta_{\rm esc}>1$. $Q_0$ is the injection rate, and the step function is $\SF{x}{a}{b}=1$ for $a\leq x\leq b$, and $0$ otherwise. Eq.~(\ref{eq:b13_kineq}) has a simple analytical solution in the form of a broken power-law:

\begin{align}
    n_e^{\rm fast}(\gamma) &= n_0 \begin{cases}
            \gamma^{-2} & \gamma_{\rm br}<\gamma\leq\gamma_{\rm min} \\
            \gamma^{-(s+1)} & \gamma_{\rm min}<\gamma\leq\gamma_{\rm max}
    \end{cases} \label{eq:b13_fast} \\
    n_e^{\rm slow}(\gamma) &= n_0 \begin{cases}
            \gamma^{-s} & \gamma_{\rm min}<\gamma\leq\gamma_{\rm br} \\
            \gamma^{-(s+1)} & \gamma_{\rm br}<\gamma\leq\gamma_{\rm max}
    \end{cases} \label{eq:b13_slow},
\end{align}
and $0$ in all other cases. The break Lorentz factor $\gamma_{\rm br}$ is derived from the equality of cooling and escape: $\gamma_{\rm br}/|\dot{\gamma}(\gamma_{\rm br})|=t_{\rm esc}$. The normalization of the electron distribution is derived from

\begin{align}
    \hat{L}_{\rm inj} = \pi R^2\Gamma^2m_ec^3 \int \td{\gamma}\gamma n_e(\gamma) 
    \label{eq:b13_norm},
\end{align}
with the injection luminosity in the observer's frame $\hat{L}_{\rm inj}$, the bulk Lorentz factor $\Gamma$, the electron rest mass $m_e$, and the speed of light $c$.

The radiation calculation is performed in the comoving frame of the emission region and susequently transformed to the observer's frame with the Doppler factor $\delta$. We assume throughout the modeling $\delta=\Gamma$.

\subsection{Electron-proton co-acceleration model} \label{sec:epcoacc}


The model introduced by \cite{Zech2021} is based on a simple stationary one-zone code combining a population of relativistic electrons
and protons in a spherical emission region of radius $R$, with a homogeneous and isotropic magnetic field of strength $B$ and Doppler factor $\delta$. Both particle distributions are initially described either by a power law with exponential cut-off, with minimum and maximum Lorentz factors $\gamma_{\rm min,p|e}$, $\gamma_{\rm max,p|e}$, and photon index $s_{e|p}=2.2$, as expected for acceleration on a mildly relativistic shock. The particle number densities are the same for both populations. The code can also describe the hardened particle distributions expected for re-acceleration on consecutive shocks, with $n_{\rm shock}$ the total number of shocks. This distribution can be
approximated by
\begin{equation}
    \frac{{\rm d}N_>^{(n)}}{{\rm d}\gamma_>}\,=\,\frac{(s-1)^{n+1}}{n!\,g^n\,\gamma_{\rm min}}\left(\frac{\gamma_>}{g^n\,\gamma_{\rm min}}\right)^{-s}\ln\left(\frac{\gamma_>}{g^n\,\gamma_{\rm min}} \right)^n\,.
    \label{eq:AD5}
\end{equation}
with $n = n_{\rm shock}$ and $g \sim 2$ in our scenario.

For electrons and protons accelerated on the same shock front, it can be shown that electrons will be preheated up to a fraction of equipartition as energy is transferred from protons to electrons.
This leads to a relation between the minimum Lorentz factors $\gamma_{min,e} \sim 600 \gamma_{min,p}$. We suppose acceleration on mildly relativistic shocks with $\gamma_{sh} \sim 3$, leading
to $\gamma_{min,p} \sim 3$ and $\gamma_{min,e} \sim 1800$. 
An additional constraint comes from the fact that particles need to be able to scatter effectively in the microturbulence upstream and downstream of the shock front to allow repeated shock crossings and thus efficient
energy gain. This leads to a relation between the minimum and maximum Lorentz factors that depends on the 
magnetization $\sigma$: $\gamma_{\rm max,e|p} \leq {\gamma_{min,e|p}} / {\sqrt{\sigma}}$.

To achieve acceptable representations of the observed SEDs within a coherent shock acceleration scenario, it is assumed that the magnetization in the emission region downstream from the shock $\sigma_{\rm rad}$ can become orders of magnitude larger than the
upstream magnetization $\sigma$. This is justified through a possible amplification of the magnetic field caused by the flow of accelerated charged particles. 

Given the low Lorentz factors of the proton population and its number density that equals that of the electrons, any radiative emission from the protons, although it is fully modelled by the
code, can be completely neglected. The presence of the proton population provides simply a physical justification for the high minimum Lorentz factors of the electrons.

\subsection{OneHaLe} \label{sec:onehale}

The \onehale\ code \citep{Zacharias21,zacharias+22} is a time-dependent, one-zone hadro-leptonic model calculating the particle distributions and photon spectra in a spherical region with radius $R$ permeated by a tangled magnetic field $B$. It moves with bulk Lorentz factor $\Gamma$, and as above we assume here $\delta=\Gamma$. The code contains various options for external fields, such as the accretion disk, the broad-line region, the dusty torus and the cosmic microwave background. However, in the application here, we do not consider the broad-line region and the dusty torus, and the accretion disk only serves as a potential contribution to the optical spectrum, but does not partake strongly in the particle-photon interactions.

The particle distribution $n_i(\chi)$ of species $i$ (protons, charged pions, muons, and electrons including positrons) is given here as a function of normalized momentum $\chi = p_i/m_i c = \gamma\beta$ with the particle mass $m_i$ and $\beta = \sqrt{1-\gamma^{-2}}$. The distributions are derived from the Fokker-Planck equation

\begin{align}
     \frac{\pd{n_i(\chi,t)}}{\pd{t}} &= \frac{\pd{}}{\pd{\chi}} \left[ \frac{\chi^2}{(a+2)t_{\rm acc}} \frac{\pd{n_i(\chi,t)}}{\pd{\chi}} \right] 
     - \frac{\pd{}}{\pd{\chi}} \left( \dot{\chi}_i n_i(\chi,t) \right) \nonumber \\
	 &\quad - \frac{n_i(\chi,t)}{t_{\rm esc}} - \frac{n_i(\chi, t)}{\gamma t^{\ast}_{i,{\rm decay}}}
	 + Q_i(\chi,t)
	 \label{eq:z21_fpgen}.
\end{align}
The first term on the right-hand-side describes Fermi-II acceleration through momentum diffusion employing hard-sphere scattering. The parameter $a$ is the ratio of shock to Alv\`{e}n speed, while $t_{\rm acc}$ is the energy-independent acceleration time scale. The second term contains continuous energy gains and losses. The gain is Fermi-I acceleration described by $\dot{\chi}_{\rm FI} = \chi/t_{\rm acc}$, while the loss term contains the radiative and adiabatic processes of each particle specie. All charged particles undergo synchrotron and adiabatic cooling, while protons additionally lose energy through Bethe-Heitler pair and pion production. We note that in the code version employed here, the conversion of protons to neutrons is not treated explicitly, but is considered as a continuous energy loss process instead. Electrons additionally undergo inverse-Compton losses scattering all available internal and external photon fields.

The third term in Eq.~(\ref{eq:z21_fpgen}) marks the escape of particles as in Sec.~\ref{sec:boettcher13}. The fourth term describes the decay of unstable particles, which decay in proper time $t^{\ast}_{i,{\rm decay}}$. The final term contains the injection of particles. Primary injection of protons and electrons follows a simple power-law as in Eq.~(\ref{eq:b13_kineq}) with normalization

\begin{align}
    Q_{0,i}(t) = \frac{L_{\rm inj,i}(t)}{\frac{4}{3}\pi R^3 m_ic^2}\left[ \intl_{\gamma_{\rm min,i}}^{\gamma_{\rm max,i}}\td{\gamma} \gamma^{1-s_i} \right]^{-1}
    \label{eq:z21_injnorm}.
\end{align}
We stress the fact that in this case the injection luminosity is given in the comoving frame, thus not including the bulk Lorentz factor as in Eq.~(\ref{eq:b13_norm}). The primary injection also includes primary acceleration indicating that the acceleration terms in Eq.~(\ref{eq:z21_fpgen}) are merely acting as a mild reacceleration with $t_{\rm acc} = \eta_{\rm acc}t_{\rm esc}$. The pion injection term directly follows from the proton-photon interactions, while muons are injected from pion decay. Secondary electrons are injected from muon decay, Bethe-Heitler pair production, and \g-\g\ pair production. Neutral pions decay quickly into \g\ rays, which is why the resulting radiation is computed directly from their injection spectrum.

As Eq.~(\ref{eq:z21_fpgen}) relies on the time-dependent photon spectrum present wihtin the emission region, in each time step we solve the Fokker-Planck equation for all charged particle species, and the radiation transport equation containing terms for photon production, absorption and escape. 

While the code is fully time-dependent, we use it here to calculate steady-state solutions. This is achieved if the total particle densities of protons $n_p$ and electrons $n_e$ each vary less than $10^{-4}$ relative to the previous two time steps. The detailed equations of the whole code can be found in \cite{zacharias+22}.
\section{Modeling parameters} \label{app:tables}

In the following tables, we list the parameters of the various models employed in Sec.~\ref{sec:modeling}. In all tables, the particle injection luminosity in the \SSC\ model is defined in the observer's frame, while it is defined in the comoving frame in the other models.

%
\begin{table*}
\centering
\footnotesize
	\caption {Parameters for 1ES~0120$+$340 for the various models, input (top), inferred jet powers (bottom).} 
\begin{tabular}{lc|cccc}
\hline \hline 
	\multicolumn{2}{l|}{Parameter} & \SSC\ & \epshock\ & \lhpi & \lhp \\
\hline 
p Inj. Luminosity & $L_{\rm inj,p}$ [erg/s] & -- & -- & $3\E{43}$ & $4\E{40}$ \\
p energy density & $u_p$ [erg/cm$^3$] & -- & $2.9\E{-4}$ & -- & -- \\
Min. p Lorentz fac. & $\gamma_{\rm min,p}$ & -- & $3$ & $2$ & $2$ \\
Max. p Lorentz fac. & $\gamma_{\rm max,p}$ & -- & $1.4\E{3}$ & $2\E{7}$ & $1\E{11}$ \\
p Inj. spectal index & $s_p$ & -- & 2.2 & 1.2 & 2.1 \\
e Inj. Luminosity & $L_{\rm inj,e}$ [erg/s] & $4.5\E{44}$ & -- & $1.6\E{39}$ & $1.5\E{39}$ \\
e energy density & $u_e$ [erg/cm$^3$] & -- & $9.7\E{-5}$ & -- & -- \\
Min. e Lorentz fac. & $\gamma_{\rm min,e}$ & $2\E{3}$ & $1.8\E{3}$ & $1\E{4}$ & $2\E{3}$ \\
Max. e Lorentz fac. & $\gamma_{\rm max,e}$ & $8\E{5}$ & $8.6\E{5}$ & $1.5\E{5}$ & $2\E{4}$ \\
e Inj. spectal index & $s_e$ & 1.2 & 2.2 & 1.2 & 1.2 \\
Radius & $R$ [cm] & $8\E{16}$ & $9.5\E{16}$ & $5\E{15}$ & $1\E{15}$ \\
Escape time par. & $\eta_{\rm esc}$ & 150 & -- & 50 & 50 \\
Acc. time par. & $\eta_{\rm acc}$ & -- & -- & 50 & 5 \\
Magnetic field & $B$ [G] & 0.01 & 0.004 & 1.0 & 10.0 \\
Doppler factor & $\delta$ & 50 & 50 & 50 & 50 \\
Number of shocks & $n_{\rm shock}$ & -- & 2 & -- & -- \\
\hline \hline 
Radiative Power & $\hat{P}_{\rm rad}$ [erg/s] & $5.4\E{42}$ & $5.6\E{42}$ & $6.5\E{42}$ & $6.3\E{42}$ \\
Magnetic Power & $\hat{P}_{\rm B}$ [erg/s]   & $1.5\E{42}$ & $2.7\E{42}$ & $2.3\E{44}$ & $9.4\E{44}$ \\
e Power & $\hat{P}_{\rm e}$ [erg/s]   & $6.5\E{43}$ & $4.0\E{44}$ & $1.8\E{42}$ & $4.6\E{41}$ \\
p Power & $\hat{P}_{\rm p}$ [erg/s]   & $4.8\E{42}$ & $1.2\E{45}$ & $1.6\E{48}$ & $2.6\E{45}$ \\
\hline \hline 
\end{tabular} \\
\label{tab:models0120}
\end{table*}
%

%
\begin{table*}
\centering
\footnotesize
	\caption {Parameters for RGB~J0710$+$591 for the various models of Spectrum 1, input (top), inferred jet powers (bottom).} 
\begin{tabular}{lc|cccc}
\hline \hline 
	\multicolumn{2}{l|}{Parameter} & \SSC\ & \epshock\ & \lhpi & \lhp \\
\hline 
p Inj. Luminosity & $L_{\rm inj,p}$ [erg/s] & -- & -- & $1\E{43}$ & $5\E{40}$ \\
p energy density & $u_p$ [erg/cm$^3$] & -- & $4.4\E{-3}$ & -- & -- \\
Min. p Lorentz fac. & $\gamma_{\rm min,p}$ & -- & 3 & $2$ & $2$ \\
Max. p Lorentz fac. & $\gamma_{\rm max,p}$ & -- & $1.4\E{3}$ & $2\E{7}$ & $4\E{10}$ \\
p Inj. spectal index & $s_p$ & -- & 2.2 & 1.5 & 1.9 \\
e Inj. Luminosity & $L_{\rm inj,e}$ [erg/s] & $1.3\E{44}$ & -- & $1\E{39}$ & $1\E{39}$ \\
e energy density & $u_e$ [erg/cm$^3$] & -- & $1.4\E{-3}$ & -- & -- \\
Min. e Lorentz fac. & $\gamma_{\rm min,e}$ & $2\E{3}$ & $1.8\E{3}$ & $1\E{4}$ & $1\E{4}$ \\
Max. e Lorentz fac. & $\gamma_{\rm max,e}$ & $1\E{6}$ & $8.6\E{5}$ & $2.5\E{5}$ & $2.5\E{5}$ \\
e Inj. spectal index & $s_e$ & 1.5 & 2.2 & 1.5 & 1.5 \\
Radius & $R$ [cm] & $1\E{16}$ & $1.5\E{16}$ & $1\E{15}$ & $1\E{15}$ \\
Escape time par. & $\eta_{\rm esc}$ & 50 & -- & 50 & 50 \\
Acc. time par. & $\eta_{\rm acc}$ & -- & -- & 50 & 5 \\
Magnetic field & $B$ [G] & 0.03 & 0.02 & 2.0 & 2.0 \\
Doppler factor & $\delta$ & 50 & 50 & 50 & 50 \\
Number of shocks & $n_{\rm shock}$ & -- & 1 & -- & -- \\
\hline \hline 
Radiative Power & $\hat{P}_{\rm rad}$ [erg/s] & $3.9\E{42}$ & $3.9\E{42}$ & $3.9\E{42}$ & $3.7\E{42}$ \\
Magnetic Power & $\hat{P}_{\rm B}$ [erg/s]   & $8.4\E{41}$ & $2.1\E{42}$ & $3.8\E{43}$ & $3.8\E{43}$ \\
e Power & $\hat{P}_{\rm e}$ [erg/s]   & $2.4\E{43}$ & $1.6\E{44}$ & $1.2\E{42}$ & $1.2\E{42}$ \\
p Power & $\hat{P}_{\rm p}$ [erg/s]   & $2.7\E{42}$ & $4.8\E{44}$ & $5.5\E{47}$ & $3.2\E{45}$ \\
\hline \hline 
\end{tabular} \\
\label{tab:models0710old1}
\end{table*}
%

%
\begin{table*}
\centering
\footnotesize
	\caption {Parameters for RGB~J0710$+$591 for the various models of Spectrum 2, input (top), inferred jet powers (bottom).} 
\begin{tabular}{lc|cccc}
\hline \hline 
	\multicolumn{2}{l|}{Parameter} & \SSC\ & \epshock\ & \lhpi & \lhp \\
\hline 
p Inj. Luminosity & $L_{\rm inj,p}$ [erg/s] & -- & -- & $2\E{43}$ & $5\E{41}$ \\
p energy density & $u_p$ [erg/cm$^3$] & -- & $4.2\E{-3}$ & -- & -- \\
Min. p Lorentz fac. & $\gamma_{\rm min,p}$ & -- & 3 & $2$ & $2$ \\
Max. p Lorentz fac. & $\gamma_{\rm max,p}$ & -- & $1.4\E{3}$ & $1\E{7}$ & $4\E{10}$ \\
p Inj. spectal index & $s_p$ & -- & 2.2 & 1.5 & 2.3 \\
e Inj. Luminosity & $L_{\rm inj,e}$ [erg/s] & $1.8\E{44}$ & -- & $3.5\E{38}$ & $3.5\E{38}$ \\
e energy density & $u_e$ [erg/cm$^3$] & -- & $1.4\E{-3}$ & -- & -- \\
Min. e Lorentz fac. & $\gamma_{\rm min,e}$ & $2\E{3}$ & $1.8\E{3}$ & $5\E{3}$ & $2\E{3}$ \\
Max. e Lorentz fac. & $\gamma_{\rm max,e}$ & $1\E{6}$ & $8.6\E{5}$ & $2\E{5}$ & $9\E{4}$ \\
e Inj. spectal index & $s_e$ & 1.5 & 2.2 & 1.5 & 1.5 \\
Radius & $R$ [cm] & $1\E{16}$ & $1.9\E{16}$ & $1\E{15}$ & $1\E{15}$ \\
Escape time par. & $\eta_{\rm esc}$ & 200 & -- & 50 & 50 \\
Acc. time par. & $\eta_{\rm acc}$ & -- & -- & 50 & 5 \\
Magnetic field & $B$ [G] & 0.02 & 0.01 & 2.0 & 10.0 \\
Doppler factor & $\delta$ & 50 & 50 & 50 & 50 \\
Number of shocks & $n_{\rm shock}$ & -- & 1 & -- & -- \\
\hline \hline 
Radiative Power & $\hat{P}_{\rm rad}$ [erg/s] & $1.6\E{42}$ & $1.7\E{42}$ & $1.6\E{42}$ & $1.4\E{42}$ \\
Magnetic Power & $\hat{P}_{\rm B}$ [erg/s]   & $3.8\E{41}$ & $6.8\E{41}$ & $3.8\E{43}$ & $9.4\E{44}$ \\
e Power & $\hat{P}_{\rm e}$ [erg/s]   & $2.6\E{43}$ & $2.5\E{44}$ & $6\E{41}$ & $8.5\E{40}$ \\
p Power & $\hat{P}_{\rm p}$ [erg/s]   & $3.7\E{42}$ & $7.6\E{44}$ & $1.1\E{48}$ & $3.3\E{46}$ \\
\hline \hline 
\end{tabular} \\
\label{tab:models0710old2}
\end{table*}
%

%
\begin{table*}
\centering
\footnotesize
	\caption {Parameters for RGB~J0710$+$591 for the various models of Spectrum 3, input (top), inferred jet powers (bottom).} 
\begin{tabular}{lc|cccc}
\hline \hline 
	\multicolumn{2}{l|}{Parameter} & \SSC\ & \epshock\ & \lhpi & \lhp \\
\hline 
p Inj. Luminosity & $L_{\rm inj,p}$ [erg/s] & -- & -- & $1\E{43}$ & $3.5\E{42}$ \\
p energy density & $u_p$ [erg/cm$^3$] & -- & $3.8\E{-3}$ & -- & -- \\
Min. p Lorentz fac. & $\gamma_{\rm min,p}$ & -- & $3$ & $2$ & $2$ \\
Max. p Lorentz fac. & $\gamma_{\rm max,p}$ & -- & $1.4\E{3}$ & $1\E{7}$ & $4\E{10}$ \\
p Inj. spectal index & $s_p$ & -- & 2.2 & 1.5 & 2.4 \\
e Inj. Luminosity & $L_{\rm inj,e}$ [erg/s] & $1.4\E{44}$ & -- & $2.5\E{38}$ & $2.3\E{38}$ \\
e energy density & $u_e$ [erg/cm$^3$] & -- & $1.2\E{-3}$ & -- & -- \\
Min. e Lorentz fac. & $\gamma_{\rm min,e}$ & $2\E{3}$ & $1.8\E{3}$ & $6\E{3}$ & $3\E{3}$ \\
Max. e Lorentz fac. & $\gamma_{\rm max,e}$ & $7\E{5}$ & $8.6\E{5}$ & $1\E{5}$ & $5\E{4}$ \\
e Inj. spectal index & $s_e$ & 1.5 & 2.2 & 1.5 & 1.5 \\
Radius & $R$ [cm] & $1\E{16}$ & $2.8\E{16}$ & $1\E{15}$ & $1\E{15}$ \\
Escape time par. & $\eta_{\rm esc}$ & 200 & -- & 50 & 50 \\
Acc. time par. & $\eta_{\rm acc}$ & -- & -- & 50 & 5 \\
Magnetic field & $B$ [G] & 0.015 & 0.005 & 2.0 & 10.0 \\
Doppler factor & $\delta$ & 50 & 50 & 50 & 50 \\
Number of shocks & $n_{\rm shock}$ & -- & 1 & -- & -- \\
\hline \hline 
Radiative Power & $\hat{P}_{\rm rad}$ [erg/s] & $1.1\E{42}$ & $1.2\E{42}$ & $1\E{42}$ & $1\E{42}$ \\
Magnetic Power & $\hat{P}_{\rm B}$ [erg/s]   & $2.1\E{41}$ & $3.6\E{41}$ & $3.8\E{43}$ & $9.4\E{44}$ \\
e Power & $\hat{P}_{\rm e}$ [erg/s]   & $2.9\E{43}$ & $4.4\E{44}$ & $5.5\E{41}$ & $6.6\E{40}$ \\
p Power & $\hat{P}_{\rm p}$ [erg/s]   & $3.5\E{42}$ & $1.4\E{45}$ & $5.6\E{47}$ & $2.4\E{47}$ \\
\hline \hline 
\end{tabular} \\
\label{tab:models0710}
\end{table*}
%

%
\begin{table*}
\centering
\footnotesize
	\caption {Parameters for 1ES~1101$-$232 for the various models, input (top), inferred jet powers (bottom).} 
\begin{tabular}{lc|cccc}
\hline \hline 
	\multicolumn{2}{l|}{Parameter} & \SSC\ & \epshock\ & \lhpi & \lhp \\
\hline 
p Inj. Luminosity & $L_{\rm inj,p}$ [erg/s] & -- & -- & $1.8\E{43}$ & $1.3\E{42}$ \\
p energy density & $u_p$ [erg/cm$^3$] & -- & $3.3\E{-4}$ & -- & -- \\
Min. p Lorentz fac. & $\gamma_{\rm min,p}$ & -- & $3$ & $2$ & $2$ \\
Max. p Lorentz fac. & $\gamma_{\rm max,p}$ & -- & $1.3\E{3}$ & $2\E{7}$ & $1\E{11}$ \\
p Inj. spectal index & $s_p$ & -- & 2.2 & 1.4 & 2.3 \\
e Inj. Luminosity & $L_{\rm inj,e}$ [erg/s] & $4.7\E{44}$ & -- & $1.5\E{39}$ & $1.3\E{39}$ \\
e energy density & $u_e$ [erg/cm$^3$] & -- & $1.1\E{-4}$ & -- & -- \\
Min. e Lorentz fac. & $\gamma_{\rm min,e}$ & $5\E{2}$ & $1.8\E{3}$ & $1\E{4}$ & $5\E{3}$ \\
Max. e Lorentz fac. & $\gamma_{\rm max,e}$ & $8\E{5}$ & $7.6\E{5}$ & $1.5\E{5}$ & $5\E{4}$ \\
e Inj. spectal index & $s_e$ & 1.4 & 2.2 & 1.4 & 1.4 \\
Radius & $R$ [cm] & $1.2\E{17}$ & $1.1\E{17}$ & $3\E{15}$ & $1\E{15}$ \\
Escape time par. & $\eta_{\rm esc}$ & 150 & -- & 50 & 50 \\
Acc. time par. & $\eta_{\rm acc}$ & -- & -- & 50 & 5 \\
Magnetic field & $B$ [G] & 0.01 & 0.003 & 1.0 & 10.0 \\
Doppler factor & $\delta$ & 50 & 50 & 50 & 50 \\
Number of shocks & $n_{\rm shock}$ & -- & 2 & -- & -- \\
\hline \hline 
Radiative Power & $\hat{P}_{\rm rad}$ [erg/s] & $5\E{42}$ & $5.6\E{42}$ & $5.7\E{42}$ & $5.4\E{42}$ \\
Magnetic Power & $\hat{P}_{\rm B}$ [erg/s]   & $2.3\E{42}$ & $2.1\E{42}$ & $8.4\E{43}$ & $9.4\E{44}$ \\
e Power & $\hat{P}_{\rm e}$ [erg/s]   & $7.1\E{43}$ & $6.4\E{44}$ & $2.7\E{42}$ & $3\E{41}$ \\
p Power & $\hat{P}_{\rm p}$ [erg/s]   & $1.5\E{43}$ & $1.9\E{45}$ & $9.8\E{47}$ & $8.7\E{46}$ \\
\hline \hline 
\end{tabular} \\
\label{tab:models1101}
\end{table*}
%

%
\begin{table*}
\centering
\footnotesize
	\caption {Parameters for 1ES~1741$+$196 for the various models of Spectrum 1, input (top), inferred jet powers (bottom).} 
\begin{tabular}{lc|cccc}
\hline \hline 
	\multicolumn{2}{l|}{Parameter} & \SSC\ & \epshock\ & \lhpi & \lhp \\
\hline 
p Inj. Luminosity & $L_{\rm inj,p}$ [erg/s] & -- & -- & $1.1\E{44}$ & $1.2\E{44}$ \\
Min. p Lorentz fac. & $\gamma_{\rm min,p}$ & -- & -- & $2$ & $2$ \\
Max. p Lorentz fac. & $\gamma_{\rm max,p}$ & -- & -- & $4.5\E{7}$ & $1\E{10}$ \\
p Inj. spectal index & $s_p$ & -- & -- & 1.8 & 2.6 \\
e Inj. Luminosity & $L_{\rm inj,e}$ [erg/s] & $6\E{43}$ & -- & $1.2\E{38}$ & $7\E{37}$ \\
Min. e Lorentz fac. & $\gamma_{\rm min,e}$ & $2\E{3}$ & -- & $5\E{2}$ & $2\E{3}$ \\
Max. e Lorentz fac. & $\gamma_{\rm max,e}$ & $4\E{5}$ & -- & $4\E{5}$ & $1\E{5}$ \\
e Inj. spectal index & $s_e$ & 1.8 & -- & 1.8 & 1.8 \\
Radius & $R$ [cm] & $1.1\E{15}$ & -- & $3\E{15}$ & $1\E{15}$ \\
Escape time par. & $\eta_{\rm esc}$ & 400 & -- & 100 & 50 \\
Acc. time par. & $\eta_{\rm acc}$ & -- & -- & 50 & 5 \\
Magnetic field & $B$ [G] & 0.1 & -- & 0.1 & 10.0 \\
Doppler factor & $\delta$ & 50 & -- & 50 & 50 \\
\hline \hline 
Radiative Power & $\hat{P}_{\rm rad}$ [erg/s] & $3.7\E{41}$ & -- & $5.2\E{41}$ & $3.5\E{41}$ \\
Magnetic Power & $\hat{P}_{\rm B}$ [erg/s]   & $1.1\E{41}$ & -- & $8.4\E{41}$ & $9.4\E{44}$ \\
e Power & $\hat{P}_{\rm e}$ [erg/s]   & $9.6\E{42}$ & -- & $5.4\E{42}$ & $2.2\E{40}$ \\
p Power & $\hat{P}_{\rm p}$ [erg/s]   & $3.6\E{42}$ & -- & $9.0\E{48}$ & $8.3\E{48}$ \\
\hline \hline 
\end{tabular} \\
\label{tab:models1741old}
\end{table*}
%

%
\begin{table*}
\centering
\footnotesize
	\caption {Parameters for 1ES~1741$+$196 for the various models of Spectrum 2, input (top), inferred jet powers (bottom).} 
\begin{tabular}{lc|cccc}
\hline \hline 
	\multicolumn{2}{l|}{Parameter} & \SSC\ & \epshock\ & \lhpi & \lhp \\
\hline 
p Inj. Luminosity & $L_{\rm inj,p}$ [erg/s] & -- & -- & $6\E{43}$ & $1\E{44}$ \\
Min. p Lorentz fac. & $\gamma_{\rm min,p}$ & -- & -- & $2$ & $2$ \\
Max. p Lorentz fac. & $\gamma_{\rm max,p}$ & -- & -- & $2\E{7}$ & $1\E{10}$ \\
p Inj. spectal index & $s_p$ & -- & -- & 1.4 & 2.6 \\
e Inj. Luminosity & $L_{\rm inj,e}$ [erg/s] & $9\E{43}$ & -- & $2\E{38}$ & $1.5\E{38}$ \\
Min. e Lorentz fac. & $\gamma_{\rm min,e}$ & $5\E{2}$ & -- & $2\E{3}$ & $6\E{3}$ \\
Max. e Lorentz fac. & $\gamma_{\rm max,e}$ & $3\E{5}$ & -- & $4\E{5}$ & $6\E{4}$ \\
e Inj. spectal index & $s_e$ & 1.4 & -- & 1.4 & 1.4 \\
Radius & $R$ [cm] & $1.3\E{15}$ & -- & $3\E{15}$ & $1\E{15}$ \\
Escape time par. & $\eta_{\rm esc}$ & 400 & -- & 50 & 50 \\
Acc. time par. & $\eta_{\rm acc}$ & -- & -- & 50 & 5 \\
Magnetic field & $B$ [G] & 0.15 & -- & 0.2 & 10.0 \\
Doppler factor & $\delta$ & 50 & -- & 50 & 50 \\
\hline \hline 
Radiative Power & $\hat{P}_{\rm rad}$ [erg/s] & $6.3\E{41}$ & -- & $8.5\E{41}$ & $6\E{41}$ \\
Magnetic Power & $\hat{P}_{\rm B}$ [erg/s]   & $3.6\E{41}$ & -- & $3.4\E{42}$ & $9.4\E{44}$ \\
e Power & $\hat{P}_{\rm e}$ [erg/s]   & $6.4\E{42}$ & -- & $2.6\E{42}$ & $2.9\E{40}$ \\
p Power & $\hat{P}_{\rm p}$ [erg/s]   & $5.1\E{42}$ & -- & $3.3\E{48}$ & $6.9\E{48}$ \\
\hline \hline 
\end{tabular} \\
\label{tab:models1741a}
\end{table*}
%

%
\begin{table*}
\centering
\footnotesize
	\caption {Parameters for 1ES~1741$+$196 for the various models of Spectrum 3, input (top), inferred jet powers (bottom).} 
\begin{tabular}{lc|cccc}
\hline \hline 
	\multicolumn{2}{l|}{Parameter} & \SSC\ & \epshock\ & \lhpi & \lhp \\
\hline 
p Inj. Luminosity & $L_{\rm inj,p}$ [erg/s] & -- & -- & $8\E{43}$ & $1\E{44}$ \\
Min. p Lorentz fac. & $\gamma_{\rm min,p}$ & -- & -- & $2$ & $2$ \\
Max. p Lorentz fac. & $\gamma_{\rm max,p}$ & -- & -- & $3\E{7}$ & $1\E{10}$ \\
p Inj. spectal index & $s_p$ & -- & -- & 1.4 & 2.6 \\
e Inj. Luminosity & $L_{\rm inj,e}$ [erg/s] & $8\E{43}$ & -- & $1.2\E{38}$ & $1.1\E{38}$ \\
Min. e Lorentz fac. & $\gamma_{\rm min,e}$ & $5\E{2}$ & -- & $2\E{3}$ & $5\E{3}$ \\
Max. e Lorentz fac. & $\gamma_{\rm max,e}$ & $3\E{5}$ & -- & $3\E{5}$ & $6\E{4}$ \\
e Inj. spectal index & $s_e$ & 1.4 & -- & 1.4 & 1.4 \\
Radius & $R$ [cm] & $1.5\E{15}$ & -- & $3\E{15}$ & $1\E{15}$ \\
Escape time par. & $\eta_{\rm esc}$ & 400 & -- & 50 & 50 \\
Acc. time par. & $\eta_{\rm acc}$ & -- & -- & 50 & 5 \\
Magnetic field & $B$ [G] & 0.1 & -- & 0.2 & 10.0 \\
Doppler factor & $\delta$ & 50 & -- & 50 & 50 \\
\hline \hline 
Radiative Power & $\hat{P}_{\rm rad}$ [erg/s] & $4.7\E{41}$ & -- & $6.5\E{41}$ & $4.7\E{41}$ \\
Magnetic Power & $\hat{P}_{\rm B}$ [erg/s]   & $2.1\E{41}$ & -- & $3.4\E{42}$ & $9.4\E{44}$ \\
e Power & $\hat{P}_{\rm e}$ [erg/s]   & $7.4\E{42}$ & -- & $1.8\E{42}$ & $2.3\E{40}$ \\
p Power & $\hat{P}_{\rm p}$ [erg/s]   & $4.5\E{42}$ & -- & $4.4\E{48}$ & $6.9\E{48}$ \\
\hline \hline 
\end{tabular} \\
\label{tab:models1741b}
\end{table*}
%

%
\begin{table*}
\centering
\footnotesize
	\caption {Parameters for 1ES~2322$-$409 for the various models, input (top), inferred jet powers (bottom).} 
\begin{tabular}{lc|cccc}
\hline \hline 
	\multicolumn{2}{l|}{Parameter} & \SSC\ & \epshock\ & \lhpi & \lhp \\
\hline 
p Inj. Luminosity & $L_{\rm inj,p}$ [erg/s] & -- & -- & $1\E{45}$ & $3\E{43}$ \\
Min. p Lorentz fac. & $\gamma_{\rm min,p}$ & -- & -- & $2$ & $2$ \\
Max. p Lorentz fac. & $\gamma_{\rm max,p}$ & -- & -- & $3\E{7}$ & $1\E{10}$ \\
p Inj. spectal index & $s_p$ & -- & -- & 2.3 & 2.5 \\
e Inj. Luminosity & $L_{\rm inj,e}$ [erg/s] & $2.1\E{44}$ & -- & $9\E{38}$ & $8\E{38}$ \\
Min. e Lorentz fac. & $\gamma_{\rm min,e}$ & $4\E{3}$ & -- & $3\E{3}$ & $1.1\E{3}$ \\
Max. e Lorentz fac. & $\gamma_{\rm max,e}$ & $9\E{5}$ & -- & $2\E{5}$ & $5\E{4}$ \\
e Inj. spectal index & $s_e$ & 2.5 & -- & 2.5 & 2.5 \\
Radius & $R$ [cm] & $7\E{16}$ & -- & $2\E{15}$ & $4\E{15}$ \\
Escape time par. & $\eta_{\rm esc}$ & 40 & -- & 50 & 50 \\
Acc. time par. & $\eta_{\rm acc}$ & -- & -- & 50 & 5 \\
Magnetic field & $B$ [G] & 0.012 & -- & 1.0 & 10.0 \\
Doppler factor & $\delta$ & 50 & -- & 50 & 50 \\
\hline \hline 
Radiative Power & $\hat{P}_{\rm rad}$ [erg/s] & $3.7\E{42}$ & -- & $3.4\E{42}$ & $3.4\E{42}$ \\
Magnetic Power & $\hat{P}_{\rm B}$ [erg/s]   & $6.6\E{42}$ & -- & $3.8\E{43}$ & $1.5\E{46}$ \\
e Power & $\hat{P}_{\rm e}$ [erg/s]   & $9\E{43}$ & -- & $7.2\E{42}$ & $2.6\E{41}$ \\
p Power & $\hat{P}_{\rm p}$ [erg/s]   & $1.7\E{43}$ & -- & $6.1\E{49}$ & $2.1\E{48}$ \\
\hline \hline 
\end{tabular} \\
\label{tab:models2322}
\end{table*}
%
\end{appendix}

\end{document}